\documentclass[reprint,aps,pre,superscriptaddress]{revtex4-1}

\usepackage[utf8]{inputenc}
\usepackage[english]{babel}

\usepackage{amsmath} 
\usepackage{amssymb} 
\usepackage{physics} 
\usepackage{mathtools} 

\usepackage{graphicx} 
\graphicspath{ {./images/} } 
\usepackage[dvipsnames]{xcolor} 
\usepackage{tikz} 
    \usetikzlibrary{arrows.meta}
    \usetikzlibrary{decorations.pathreplacing}
    \usetikzlibrary{shapes.geometric}

\usepackage{array} 
\usepackage{dcolumn} 

\usepackage{hyperref} 

\makeatletter
\renewcommand{\fnum@figure}{FIG. \thefigure} 
\makeatother

\definecolor{MyRed}{rgb}{0.8, 0.2, 0.2}
\definecolor{MyBlue}{rgb}{0.0, 0.3, 0.7}
\definecolor{MyGreen}{rgb}{0.2, 0.7, 0.2}
\definecolor{MyYellow}{rgb}{1.0, 0.8, 0.0}
\definecolor{MyPurple}{rgb}{0.6, 0.0, 0.6}
\definecolor{MyOrange}{rgb}{1.0, 0.6, 0.0}
\definecolor{VacuaOrange}{rgb}{1.0, 0.8, 0.5}
\definecolor{VacuaBlueLight}{rgb}{0.6, 0.8, 1.0}
\definecolor{VacuaBlue}{rgb}{0.4, 0.7, 0.9}
\definecolor{VacuaPurpleLight}{rgb}{0.9, 0.7, 0.9}
\definecolor{VacuaPurple}{rgb}{0.8, 0.5, 0.8}
\definecolor{QuasiparticleGreen}{rgb}{0.4, 0.8, 0.4}
\definecolor{QuasiparticleRed}{rgb}{0.9, 0.3, 0.4}
\definecolor{QuasiparticleYellow}{rgb}{1.0, 0.8, 0.4}

\renewcommand{\bm}[1]{\mathbf{#1}} 
\newcommand{\um}[1]{\underline{#1}} 
\newcommand{\n}{\um{n}} 
\newcommand{\q}{Q_{\um{n}}} 
\newcommand{\m}[1]{\mathcal{M}_{\text{#1}}} 
\newcommand{\e}{\text{e}} 
\newcommand{\Mo}{\mathbf{M}_{\text{O}}}
\newcommand{\Me}{\mathbf{M}_{\text{E}}}
\newcommand{\U}{\mathbf{U}}
\newcommand{\LL}{\mathbf{L}}
\newcommand{\RR}{\mathbf{R}}
\newcommand{\vW}{\mathbf{V}^{\phantom{\prime}}}
\newcommand{\vV}{\mathbf{V}^{\prime}}
\newcommand{\vWV}{\mathbf{V}^{(\prime)}}
\newcommand{\W}{V^{\phantom{\prime}}}
\newcommand{\V}{V^{\prime}}
\newcommand{\WV}{V^{(\prime)}}
\newcommand{\tW}{\tilde{V}^{\phantom{\prime}}}
\newcommand{\tV}{\tilde{V}^{\prime}}
\newcommand{\tWV}{\tilde{V}^{(\prime)}}

\newcommand{\muN}{\mu^{-}}
\newcommand{\muP}{\mu^{+}}

\begin{document}

\preprint{APS/123-QED}

\title{Exact solution of the Floquet-PXP cellular automaton}

\author{Joseph W. P. Wilkinson} 
\thanks{These authors contributed equally}
\affiliation{School of Physics and Astronomy, University of Nottingham, Nottingham, NG7 2RD, United Kingdom} 
\affiliation{Centre for the Mathematics and Theoretical Physics of Quantum Non-equilibrium Systems, University of Nottingham, Nottingham, NG7 2RD, United Kingdom} 

\author{Katja Klobas} 
\thanks{These authors contributed equally}
\affiliation{Department of Physics, Faculty of Mathematics and Physics, University of Ljubljana, Ljubljana, Slovenia} 

\author{Toma\v{z} Prosen} 
\affiliation{Department of Physics, Faculty of Mathematics and Physics, University of Ljubljana, Ljubljana, Slovenia} 

\author{Juan P. Garrahan} 
\affiliation{School of Physics and Astronomy, University of Nottingham, Nottingham, NG7 2RD, United Kingdom} 
\affiliation{Centre for the Mathematics and Theoretical Physics of Quantum Non-equilibrium Systems, University of Nottingham, Nottingham, NG7 2RD, United Kingdom} 

\date{\today} 

\begin{abstract} 
    We study the dynamics of a bulk deterministic Floquet model, the Rule 201 synchronous one-dimensional reversible cellular automaton (RCA201). The system corresponds to a deterministic, reversible, and discrete version of the PXP model, whereby a site flips only if both its nearest neighbours are unexcited. We show that the RCA201/Floquet-PXP model exhibits ballistic propagation of interacting quasiparticles - or solitons - corresponding to the domain walls between non-trivial three-fold vacuum states. Starting from the quasiparticle picture, we find the exact matrix product state form of the non-equilibrium stationary state for a range of boundary conditions, including both periodic and stochastic. We discuss further implications of the integrability of the model.
\end{abstract}


\maketitle


\section{\label{sec:introduction} Introduction}
In this paper we study the dynamics of a deterministic reversible cellular automaton (RCA), the rule 201 RCA in the classification of~\cite{Bobenko1993} or alternatively the ``Floquet-PXP'' model (named so for reasons explained below). This is a lattice system with dynamics subject to a local kinetic constraint, whose evolution is defined in terms of a local update rule which can be coded in terms of a periodic circuit, and that we show to be exactly solvable. We do this by constructing an algebraic cancellation structure which demonstrates the model's integrability. This is therefore a problem that relates to three distinct areas of current research in condensed matter theory and statistical mechanics, namely, constrained dynamics, ``Floquet'' systems, and integrability. 

Constrained systems are of interest because they often display rich collective behaviour, most notably in their dynamics. Such systems have explicit constraints either in the definition of their state spaces or in their dynamical rules. A typical example of the latter class are fully-packed dimer coverings of a lattice~\cite{Fisher1961,Henley2010,Moessner2011,Chalker2014} where only certain configurations are allowed (those with no-overlapping dimers and no uncovered sites). Among the former class are kinetically constrained models (KCMs)~\cite{Fredrickson1984,Palmer1984,Jackle1991,Ritort2003,Garrahan2011}, systems where dynamical rules are such that configurational changes can only occur if a certain local condition - the kinetic constraint - is satisfied. KCMs were originally introduced to model the slow cooperative dynamics of classical glasses (see e.g.~\cite{Ritort2003,Garrahan2011,Garrahan2018} for reviews). More recently they have been generalised to address questions in quantum non-equilibrium physics, including slow relaxation in the absence of disorder~\cite{Horssen2015,Lan2018}, as an effective description of strongly interacting Rydberg atoms~\cite{Lesanovsky2011}, and as systems displaying non-thermal eigenstates~\cite{Turner2018,Pancotti2020}. 

In systems like dimer coverings, transitions are only possible within the constrained space of states, implying constraints in the dynamics. Conversely, if in a KCM the kinetic constraint is strong enough, a configurational subspace may become dynamically disconnected thus becoming in effect a system with a constrained state space. The RCA201/Floquet-PXP model we consider here is of this kind: dynamical rules imply the existence of certain locally conserved quantities, breaking the state space into constrained subspaces disconnected by the dynamics. (In stochastic systems this is referred to as reducibility of the dynamics~\cite{Ritort2003}, a concept distinct from non-ergodicity which corresponds to the inability to forget initial conditions in finite time within a connected component.)

The second area of interest that our paper connects to are (brick-wall like) circuit systems. By this we mean systems with space-time discrete dynamics defined in terms of local gates applied synchronously throughout the system. The set of all of these gates in space and over time forms the ``circuit''. This has become a much studied problem in quantum many-body physics, where the gates correspond to unitary (or unitary and dissipative) transformations. Quantum circuits provide tractable models to study questions of entanglement, chaos, operator spreading and localisation~\cite{Nahum1,Nahum2,DeLuca,Bertini2019,Pollmann,Pollman2,sunderhauf,Khemani,Pretko}. Furthermore, when the sequence of applied gates is repeated periodically we refer to those as Floquet systems. The circuit platform is not only useful in unitary quantum many-body framework, but also in classical deterministic systems of continuous~\cite{Krajnik2020} or discrete variables (RCAs)~\cite{Klobas2020}. Moreover, so-called duality symmetries under the swap of space and time axes allow for remarkable advancements in analytic tractability~\cite{Bertini2019,Krajnik2020,Klobas2020}.

Classically, the prototypical circuit models are cellular automata (CA)~\cite{Wolfram1983,Ilachinski2001}. CAs can be both deterministic and stochastic. If deterministic, they can either be reversible or not, where the former (RCA~\cite{Bobenko1993}, see also~\cite{Takesue1987}) can be considered as a model of classical many-body Hamiltonian (or symplectic) dynamics. The RCA201/Floquet-PXP is a deterministic RCA, closely related to the now much studied RCA54/Floquet-FA~\cite{Prosen2016,Inoue2018,Prosen2017,Buca2019,Friedman2019,Gopalakrishnan2018,Gopalakrishnan2018b,Klobas2019,Klobas2019b,Alba2019,Alba2020,Klobas2020}. Just like the RCA54, the RCA201 (see detailed definitions below) is a one-dimensional lattice of binary variables with local three-site gates applied simultaneously to two halves (of even/odd indexed sites) of the lattice in two successive half time-steps. The repeated application of these makes the system a Floquet one. The local gate implements the kinetic constraint in this context. In the case of RCA54, the condition for a site to flip is identical to that of the classical Fredrickson-Andersen (FA) KCM~\cite{Fredrickson1984,Ritort2003,Garrahan2018}. For this reason RCA54 is sometimes called Floquet-FA~\cite{Gopalakrishnan2018,Gopalakrishnan2018b,Friedman2019}. In the case of RCA201, the local condition for spin flips coincides with that of the PXP model~\cite{Fendley2004,Lesanovsky2011,Turner2018}. For this reason we call the RCA201 the Floquet-PXP model.

The third area to which our work here connects is that of integrable systems~\cite{korepin1997quantum,sutherland2004beautiful,baxter2016exactly}. In particular, the RCA54/Floquet-FA was shown to be integrable~\cite{Bobenko1993,Prosen2016}, with elementary excitations corresponding to interacting localized quasiparticles (also referred to as {\em solitons} in our context). From this observation many results followed: the exact matrix product state (MPS) form of the steady state distribution in the presence of stochastic reservoirs~\cite{Prosen2016,Inoue2018}, the dominant decay modes~\cite{Prosen2017}, the exact large deviation statistics of dynamical observables~\cite{Buca2019}, the explicit MPS representation of the complete time evolution of local observables~\cite{Klobas2019}, and the exact MPS representation of multi-time correlations~\cite{Klobas2019b}. In this sense, the RCA54 is essentially a completely solved model, despite the fact that a highly versatile cubic algebraic cancellation mechanism put forward in~\cite{Prosen2017} has not (yet) been related to more standard Yang-Baxter integrability structures. Here we show that the RCA201/Floquet-PXP is also integrable in the same sense as RCA54 and propose the corresponding algebratic cancellation scheme. There is however a remarkable difference, namely RCA201 has a topological structure of muliple vacua, and quasiparticles (connecting distinct vaccuum states) which interact attractively (rather than repulsively as in the RCA54). As for the RCA54, our construction allows us to obtain a number of results for RCA201/Floquet-PXP, like the exact MPS solution of its non-equilibrium stationary state (NESS) in a range of boundary conditions that we present here. 

\bigskip

\noindent
{\bf Note added:} Upon completion of this work we became aware of the very recent Ref.~\cite{Iadecola2020} which also considers the RCA201/Floquet-PXP model. While focusing mostly on its quantum generalisation, Ref.~\cite{Iadecola2020} makes several observations about the classical RCA201/Floquet-PXP model, notably its integrability due the conserved quasiparticles, that coincide with the ones we make also here (we refer the reader specifically to Appendix A of Ref.~\cite{Iadecola2020}). In our paper here, however, we prove exactly these and various other results.

\bigskip

The paper is organised as follows. In Sec.~\ref{sec:model} we introduce the model, discuss its kinematics and basic dynamics, in particular the definition of conserved quasiparticles. In Sec.~\ref{sec:periodic} we consider dynamics under periodic boundary conditions, that is, when evolution is completely deterministic. The main result of that section is the exact NESS, in the form of a Gibbs state of the density of solitons represented as an MPS. In Sec.~\ref{sec:stochastic} we consider the case of stochastic boundaries, which can be obtained as a reduction of the periodic boundary case, and compute the exact MPS form of the corresponding NESS. In Sec.~\ref{sec:conclusion} we provide our conclusion and an outlook of future work.

\section{\label{sec:model} Floquet-PXP model}
\subsection{\label{sec:dynamics} Definition of the dynamics}

We consider a system defined on a chain of even size $N$ of binary variables $n_{i} \in \{0, 1\}$ on sites $i \in \{1, \ldots, N\}$. At discrete time $t$ the system is characterized by a configuration represented by a binary string,
\begin{equation}
    \n^{t} \equiv (n_{1}^{t}, n_{2}^{t}, \ldots, n_{N}^{t}) \in \{0, 1\}^{\times N}.
\end{equation}
The site $i$ at time $t$ is referred to as empty (or down) if $n_{i}^{t} = 0$ and occupied (or up) if $n_{i}^{t} = 1$. The dynamics of the system is given by the staggered discrete space-time mapping
\begin{equation}\label{eq:even-odd-maps}
    \n^{t + 1} =
    \begin{cases}
        \m{E}(\n^{t}), & t = 0 \pmod{2}, \\
        \m{O}(\n^{t}), & t = 1 \pmod{2}, \\
    \end{cases}
\end{equation}
where $\m{E}$ and $\m{O}$ are maps defined by local updates, 
\begin{equation}
    n_{i}^{t + 1} =
    \begin{cases}
        f_{i}^{t}, & i + t = 0 \pmod{2}, \\
        n_{i}^{t}, & i + t = 1 \pmod{2}, \\
    \end{cases}
\end{equation}
with
\begin{equation}\label{eq:local-update}
    f_{i}^{t} \equiv f(n_{i - 1}^{t}, n_{i}^{t}, n_{i + 1}^{t}),
\end{equation}
denoting a local three-site update rule (or ``gate'') acting on site $i$.

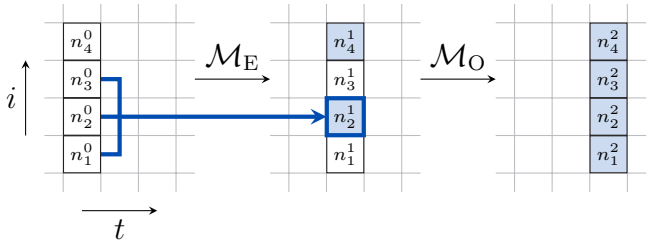
\begin{figure}[htp]
    \centering
    \vspace{10pt}
    \begin{tikzpicture}[
        scale=0.5,
        l/.style={draw=Black!30}, 
        s/.style={draw=Black, inner sep=0pt, minimum size=14pt}, 
        u/.style={fill=MyBlue!20}, 
        r/.style={ultra thick, >=stealth, draw=MyBlue}, 
        ]
        \draw[l] (-0.5,-0.5) grid (3.5,4.5);
        \draw[l] (5.5,-0.5) grid (9.5,4.5);
        \draw[l] (11.5,-0.5) grid (15.5,4.5);
        \draw[r] (1,0.5) -| (1.5,1.5);
        \draw[r, ->, >=stealth, MyBlue] (1,1.5) -- (7,1.5);
        \draw[r] (1,2.5) -| (1.5,1.5);
        \node[s] at (0.5,0.5) {\scriptsize $n_{1}^{0}$};
        \node[s] at (0.5,1.5) {\scriptsize $n_{2}^{0}$};
        \node[s] at (0.5,2.5) {\scriptsize $n_{3}^{0}$};
        \node[s] at (0.5,3.5) {\scriptsize $n_{4}^{0}$};
        \node[s] at (7.5,0.5) {\scriptsize $n_{1}^{1}$};
        \node[s] at (7.5,2.5) {\scriptsize $n_{3}^{1}$};
        \node[s, u, r] at (7.5,1.5) {\scriptsize $n_{2}^{1}$};
        \node[s, u] at (7.5,3.5) {\scriptsize $n_{4}^{1}$};
        \node[s, u] at (14.5,0.5) {\scriptsize $n_{1}^{2}$};
        \node[s, u] at (14.5,1.5) {\scriptsize $n_{2}^{2}$};
        \node[s, u] at (14.5,2.5) {\scriptsize $n_{3}^{2}$};
        \node[s, u] at (14.5,3.5) {\scriptsize $n_{4}^{2}$};
        \draw[->, >=stealth] (3.5,2.5) -- node[above] {\large $\m{E}$} (5.5,2.5);
        \draw[->, >=stealth] (9.5,2.5) -- node[above] {\large $\m{O}$} (11.5,2.5);
        \draw[->, >=stealth] (-1,1) -- node[left] {\large $i$} (-1,3);
        \draw[->, >=stealth] (0.5,-1) -- node[below] {\large $t$} (2.5,-1);
    \end{tikzpicture}
    \caption{\textbf{Dynamical scheme.} Evolution of four sites of the lattice under one full time-step of the dynamics. Shaded squares indicate which sites are updated by the local gates under each half time-step of the dynamics.}
    \label{fig:evolution}
\end{figure}

One full step of time evolution is given by the successive application of the even and odd maps, $\m{E}$ and $\m{O}$, respectively, see Eq.~\eqref{eq:even-odd-maps}, \begin{equation}\label{eq:map}
    \m{}(\n^{t}) \equiv \m{O} \big(\m{E}(\n^{t})\big),\quad
    \m{}=\m{O}\circ\m{E}.
\end{equation}
As the map $\m{}$ is applied periodically, we call this a Floquet dynamics. A schematic representation of the discrete time evolution \eqref{eq:map} is presented in Fig.~\ref{fig:evolution}.

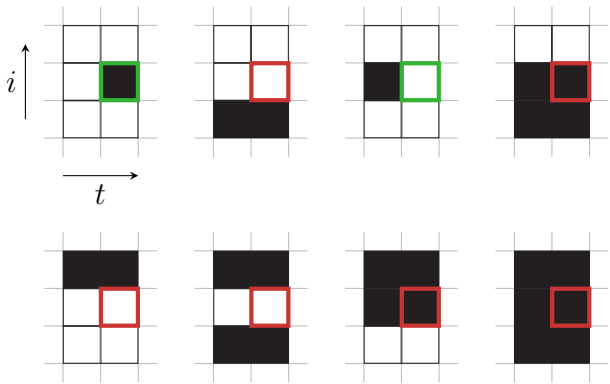
\begin{figure}[htp]
    \centering
    \vspace{10pt}
    \begin{tikzpicture}[
        scale=0.5,
        l/.style={draw=Black!30},
        e/.style={draw=Black, fill=White, minimum size=14pt},
        o/.style={draw=Black, fill=Black, minimum size=14pt},
        g/.style={ultra thick, draw=MyGreen},
        r/.style={ultra thick, draw=MyRed},
        ]
        \draw[l] (-0.5,-0.5) grid (2.5,3.5);
        \draw[l] (3.5,-0.5) grid (6.5,3.5);
        \draw[l] (7.5,-0.5) grid (10.5,3.5);
        \draw[l] (11.5,-0.5) grid (14.5,3.5);
        \draw[l] (-0.5,-6.5) grid (2.5,-2.5);
        \draw[l] (3.5,-6.5) grid (6.5,-2.5);
        \draw[l] (7.5,-6.5) grid (10.5,-2.5);
        \draw[l] (11.5,-6.5) grid (14.5,-2.5);
        \node[e] at (0.5,0.5) {};
        \node[e] at (0.5,1.5) {};
        \node[e] at (0.5,2.5) {};
        \node[e] at (1.5,0.5) {};
        \node[e] at (1.5,2.5) {};
        \node[o, g] at (1.5,1.5) {};
        \node[o] at (4.5,0.5) {};
        \node[e] at (4.5,1.5) {};
        \node[e] at (4.5,2.5) {};
        \node[o] at (5.5,0.5) {};
        \node[e] at (5.5,2.5) {};
        \node[e, r] at (5.5,1.5) {};
        \node[e] at (8.5,0.5) {};
        \node[o] at (8.5,1.5) {};
        \node[e] at (8.5,2.5) {};
        \node[e] at (9.5,0.5) {};
        \node[e] at (9.5,2.5) {};
        \node[e, g] at (9.5,1.5) {};
        \node[o] at (12.5,0.5) {};
        \node[o] at (12.5,1.5) {};
        \node[e] at (12.5,2.5) {};
        \node[o] at (13.5,0.5) {};
        \node[e] at (13.5,2.5) {};
        \node[o, r] at (13.5,1.5) {};
        \node[e] at (0.5,-5.5) {};
        \node[e] at (0.5,-4.5) {};
        \node[o] at (0.5,-3.5) {};
        \node[e] at (1.5,-5.5) {};
        \node[o] at (1.5,-3.5) {};
        \node[e, r] at (1.5,-4.5) {};
        \node[o] at (4.5,-5.5) {};
        \node[e] at (4.5,-4.5) {};
        \node[o] at (4.5,-3.5) {};
        \node[o] at (5.5,-5.5) {};
        \node[o] at (5.5,-3.5) {};
        \node[e, r] at (5.5,-4.5) {};
        \node[e] at (8.5,-5.5) {};
        \node[o] at (8.5,-4.5) {};
        \node[o] at (8.5,-3.5) {};
        \node[e] at (9.5,-5.5) {};
        \node[o] at (9.5,-3.5) {};
        \node[o, r] at (9.5,-4.5) {};
        \node[o] at (12.5,-5.5) {};
        \node[o] at (12.5,-4.5) {};
        \node[o] at (12.5,-3.5) {};
        \node[o] at (13.5,-5.5) {};
        \node[o] at (13.5,-3.5) {};
        \node[o, r] at (13.5,-4.5) {};
        \draw[->, >=stealth] (-1,0.5) -- node[left] {\large $i$} (-1,2.5);
        \draw[->, >=stealth] (0,-1) -- node[below] {\large $t$} (2,-1);
    \end{tikzpicture}
    \caption{\textbf{Rule 201.} Illustration of the action of the local gates implementing RCA201 evolution, as defined in Eq.~\eqref{eq:rule-201}. White and black squares represent empty and occupied sites, respectively. The vertical direction is space, and the horizontal is time. In each of the diagrams, only the central site is updated. The green and red borders highlight whether the site has changed or not under the action of the gate.}
    \label{fig:rule-201}
\end{figure}

In the bulk, $i \in \{2, \ldots, N - 1\}$, the discrete dynamics is given by the deterministic RCA rule 201 (RCA201) function \cite{Bobenko1993},
\begin{equation}\label{eq:rule-201}
    f_{i}^{t} = 1 + n_{i - 1}^{t} + n_{i}^{t} + n_{i + 1}^{t} + n_{i - 1}^{t} n_{i + 1}^{t} \pmod{2}.
\end{equation}
A diagrammatic illustration of the local update rule is depicted in Fig.~\ref{fig:rule-201}. This update rule can be thought of as a kinetic constraint: site $i$ can only flip if both its nearest neighbours are unexcited (and it does so deterministically). In the KCM jargon it corresponds to the constraint of the ``two-spin facilitated'' Fredrickson-Andersen model \cite{Ritort2003}. This constraint is the same as that of the kinetic energy in the PXP model \cite{Fendley2004,Lesanovsky2011,Turner2018}, and from it follows the alternative name of the RCA201 model. 

Here and in the next section we will assume that the whole system is closed, of even size $N$, and has periodic boundary conditions (PBCs). In later sections we generalise to other kinds of boundaries. PBCs are imposed in the usual manner by identifying a pair of sites $n_{0}^{t} \equiv n_{N}^{t}$ and $n_{N+1}^{t} \equiv n_{1}^{t}$. The dynamics for the sites at the left and right boundaries, $i \in \{1, N\}$, is then given by boundary functions equivalent to the RCA201 function~\eqref{eq:rule-201}, 
\begin{equation}\label{eq:boundary-update}
\begin{aligned}
    f_{1}^{t} & \equiv f(n_{N}^{t}, n_{1}^{t}, n_{2}^{t}), \\
    f_{N}^{t} & \equiv f(n_{N - 1}^{t}, n_{N}^{t}, n_{1}^{t}).
\end{aligned}
\end{equation}

\subsection{\label{sec:space} Structure of the configuration space}

\begin{figure}[htp]
    \vspace{10pt}
    \centering
    \begin{tikzpicture}
        \node[very thick, draw=Black, inner sep=0] at (0,-7.6) {\includegraphics[trim={34pt 87pt 34pt 89pt}, clip, width=.42\textwidth]{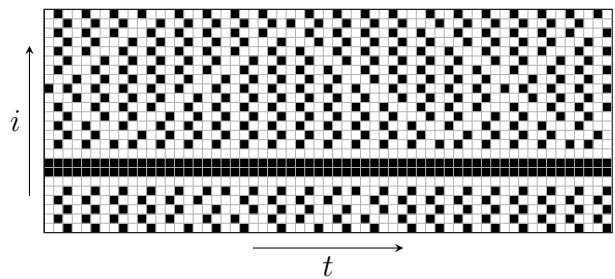}};
        \draw[->, >=stealth] (-3.97,-8.6) -- node[left] {\large $i$} (-3.97,-6.6);
        \draw[->, >=stealth] (-1,-9.29) -- node[below] {\large $t$} (1,-9.29);
    \end{tikzpicture}
    \caption{\textbf{RCA201/Floquet-PXP trajectory.} 
    A trajectory of the model with PBC illustrating the spatial localization of pairs of excited sites. In this trajectory there are two solitons that change 
    direction under reflection with the localised pair. Note also the distinct cycles of the vacua motifs. In the rest of the paper we focus on the configurational sector with no pairs of excited neighbours.}
    \label{fig:site-invariance}
\end{figure}

The local dynamics generated by the RCA201 function~\eqref{eq:rule-201} imposes a constraint on the system that derives from the spatial localization (immobility) of adjacent occupied sites within configurations, $\n = (\ldots, 1, 1, \ldots)$. Such pairs of excited sites are invariant under time evolution, as illustrated in Fig.~\ref{fig:site-invariance}. The kinetic constraint therefore makes the set of configurations $\mathbb{N} = \{0,1\}^{\times N}$ {\em reducible} under the dynamics, that is, it becomes partitioned into disjoint subsets, or irreducible components, spanned by distinct subsets of dynamically connected configurations identified by the positions of pairs of adjacent occupied sites. The largest of these subsets, denoted by $\mathbb{N}_{0}$, contains the configuration $\n = (0, 0, \ldots, 0, 0)$ and is the unique subset of configurations that contain no adjacent occupied sites.

It is straightforward to see that the cardinality of this subset grows exponentially according to a Fibonacci-like sequence known as the \textit{Lucas sequence},
\begin{equation}\label{eq:state-space-dimension}
    |\mathbb{N}_{0}(N)| = L_{N} \sim \varphi^{N},
\end{equation}
where $L_{N}$ is the $N$\textsuperscript{th} Lucas number, defined by the recursion relation $L_{N} = L_{N - 1} + L_{N - 2}$ with $L_{1} = 1$, $L_{2} = 3$, and where $\varphi = (1 + \sqrt{5})/2$ is the golden ratio. To see this we first consider the set of configurations, denoted here by $\mathbb{N}_{0}^{\prime}$, of a \textit{non-periodic} system of size $N$ with no adjacent occupied sites. Every configuration in this system with $n_{N} = 0$ can be obtained by appending 0 to the end of every configuration of a system with $N - 1$ sites, whilst every configuration with $n_{N} = 1$ can be obtained by appending $01$ to the end of every configuration of a system with $N - 2$ sites. As such, the cardinality of the set $\mathbb{N}'_{0}$ satisfies the linear recursion relation
\begin{equation}
    \abs*{\mathbb{N}^{\prime}_{0}(N)} = \abs*{\mathbb{N}^{\prime}_{0}(N - 1)} + \abs*{\mathbb{N}^{\prime}_{0}(N - 2)},
\end{equation}
with $\abs*{\mathbb{N}^{\prime}_{0}(1)} = 2$ and $\abs*{\mathbb{N}^{\prime}_{0}(2)} = 3$. This is, of course, the celebrated Fibonacci recursion relation, and so we have
\begin{equation}\label{eq:obc-dimension}
    \abs*{\mathbb{N}^{\prime}_{0}(N)} = F_{N + 2}, \qquad N > 0,
\end{equation}
with $F_{N}$ the $N$\textsuperscript{th} Fibonacci number, defined by the relation $F_{N} = F_{N - 1} + F_{N - 2}$ with $F_{1} = 1$ and $F_{2} = 1$. 

We now impose PBC on the system which equates to eliminating all configurations with $n_{1} = n_{N} = 1$. This yields a set, denoted by $\mathbb{N}_{0}$, whose cardinality is given by
\begin{equation}
    \abs*{\mathbb{N}_{0}(N)} = \abs*{\mathbb{N}^{\prime}_{0}(N)} - \abs*{\mathbb{N}^{\prime}_{0}(N - 4)},
\end{equation}
with $\abs*{\mathbb{N}_{0}(1)} = 1$ and $\abs*{\mathbb{N}_{0}(2)} = 3$. By substituting in the result from~\eqref{eq:obc-dimension} and subsequently using the fundamental equation relating Fibonacci and Lucas numbers,
\begin{equation}
    L_{N} = F_{N + 1} + F_{N - 1},
\end{equation}
it is trivial to see that this is exactly the Lucas recursion relation provided, $\abs*{\mathbb{N}_{0}(N)} = L_{N}$, $N > 0$.

For simplicity, we shall focus the remainder of our discussion on this this subspace spanned by states with PBC whose configurations contain no adjacent occupied sites.

\subsection{\label{sec:quasiparticles} Ballistic propagation of non-trivially interacting quasiparticles}

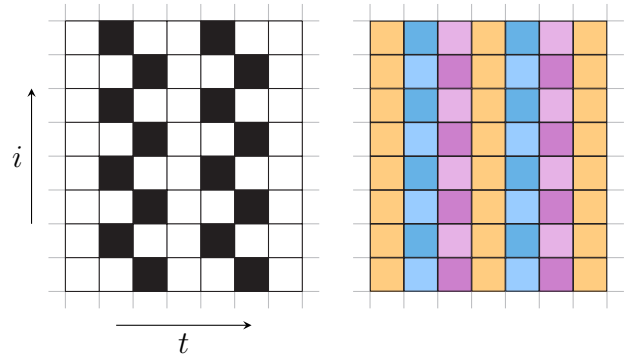
\begin{figure}[htp]
    \vspace{10pt}
    \centering
    \begin{tikzpicture}[
        scale=0.45,
        l/.style={draw=Black!30}, 
        s/.style={draw=Black, minimum size=12.6pt}, 
        e/.style={fill=White}, 
        o/.style={fill=Black}, 
        vo/.style={fill=VacuaOrange}, 
        vb/.style={fill=VacuaBlueLight}, 
        vB/.style={fill=VacuaBlue}, 
        vp/.style={fill=VacuaPurpleLight}, 
        vP/.style={fill=VacuaPurple}, 
        ]
        \draw[l] (-0.5, -0.5) grid (7.5, 8.5);
        \draw[l] (8.5, -0.5) grid (16.5, 8.5);
        \draw[s, e] (0, 0) grid (7, 8);
        \node[s, o] at (1.5, 1.5) {};
        \node[s, o] at (1.5, 3.5) {};
        \node[s, o] at (1.5, 5.5) {};
        \node[s, o] at (1.5, 7.5) {};
        \node[s, o] at (2.5, 0.5) {};
        \node[s, o] at (2.5, 2.5) {};
        \node[s, o] at (2.5, 4.5) {};
        \node[s, o] at (2.5, 6.5) {};
        \node[s, o] at (4.5, 1.5) {};
        \node[s, o] at (4.5, 3.5) {};
        \node[s, o] at (4.5, 5.5) {};
        \node[s, o] at (4.5, 7.5) {};
        \node[s, o] at (5.5, 0.5) {};
        \node[s, o] at (5.5, 2.5) {};
        \node[s, o] at (5.5, 4.5) {};
        \node[s, o] at (5.5, 6.5) {};
        \node[s, vo] at (9.5,0.5) {};
        \node[s, vo] at (9.5,1.5) {};
        \node[s, vo] at (9.5,2.5) {};
        \node[s, vo] at (9.5,3.5) {};
        \node[s, vo] at (9.5,4.5) {};
        \node[s, vo] at (9.5,5.5) {};
        \node[s, vo] at (9.5,6.5) {};
        \node[s, vo] at (9.5,7.5) {};
        \node[s, vo] at (12.5,0.5) {};
        \node[s, vo] at (12.5,1.5) {};
        \node[s, vo] at (12.5,2.5) {};
        \node[s, vo] at (12.5,3.5) {};
        \node[s, vo] at (12.5,4.5) {};
        \node[s, vo] at (12.5,5.5) {};
        \node[s, vo] at (12.5,6.5) {};
        \node[s, vo] at (12.5,7.5) {};
        \node[s, vo] at (15.5,0.5) {};
        \node[s, vo] at (15.5,1.5) {};
        \node[s, vo] at (15.5,2.5) {};
        \node[s, vo] at (15.5,3.5) {};
        \node[s, vo] at (15.5,4.5) {};
        \node[s, vo] at (15.5,5.5) {};
        \node[s, vo] at (15.5,6.5) {};
        \node[s, vo] at (15.5,7.5) {};
        \node[s, vb] at (10.5,0.5) {};
        \node[s, vB] at (10.5,1.5) {};
        \node[s, vb] at (10.5,2.5) {};
        \node[s, vB] at (10.5,3.5) {};
        \node[s, vb] at (10.5,4.5) {};
        \node[s, vB] at (10.5,5.5) {};
        \node[s, vb] at (10.5,6.5) {};
        \node[s, vB] at (10.5,7.5) {};
        \node[s, vb] at (13.5,0.5) {};
        \node[s, vB] at (13.5,1.5) {};
        \node[s, vb] at (13.5,2.5) {};
        \node[s, vB] at (13.5,3.5) {};
        \node[s, vb] at (13.5,4.5) {};
        \node[s, vB] at (13.5,5.5) {};
        \node[s, vb] at (13.5,6.5) {};
        \node[s, vB] at (13.5,7.5) {};
        \node[s, vP] at (11.5,0.5) {};
        \node[s, vp] at (11.5,1.5) {};
        \node[s, vP] at (11.5,2.5) {};
        \node[s, vp] at (11.5,3.5) {};
        \node[s, vP] at (11.5,4.5) {};
        \node[s, vp] at (11.5,5.5) {};
        \node[s, vP] at (11.5,6.5) {};
        \node[s, vp] at (11.5,7.5) {};
        \node[s, vP] at (14.5,0.5) {};
        \node[s, vp] at (14.5,1.5) {};
        \node[s, vP] at (14.5,2.5) {};
        \node[s, vp] at (14.5,3.5) {};
        \node[s, vP] at (14.5,4.5) {};
        \node[s, vp] at (14.5,5.5) {};
        \node[s, vP] at (14.5,6.5) {};
        \node[s, vp] at (14.5,7.5) {};
        \draw[->, >=stealth] (-1,2) -- node[left] {\large $i$} (-1,6);
        \draw[->, >=stealth] (1.5,-1) -- node[below] {\large $t$} (5.5,-1);
    \end{tikzpicture}
    \caption{\textbf{Vacuum configurations.} The three vacuum states are given by the spatial repetition of the motifs composed of all 0s, of alternating 0s and 1s with 1s on even sites, and alternating 0s and 1s with 1s on odd sites. In the absence of solitons, under the dynamics the three vacua repeat periodically with period three. In the panel on the right we represent the three vacuum states in orange for the all 0s, blue for the 01s, and purple for the 10s, respectively.}
    \label{fig:vacua}
\end{figure}

The physical interpretation of the dynamics in the subspace with no adjacent occupied cites, induced by the deterministic RCA201 function~\eqref{eq:rule-201}, can be intuitively understood in terms of the ballistic propagation of interacting quasiparticles representing collective excitations on a non-trivial vacuum. Specifically, the vacuum is defined as a cycle of three distinct motifs, respectively composed of repeating 0s, alternating 0s and 1s (starting and ending with 0s on odd sites), and alternating 1s and 0s (starting and ending with 0s on even sites), as illustrated in Fig.~\ref{fig:vacua}. Indeed, it can be easily demonstrated that the configurations composed entirely of repeating these three distinct arrangements form a unique, invariant trajectory, which we call a \textit{vacuum} trajectory,
\begin{equation}\label{eq:vacuum-trajectory}
\begin{aligned}
    & (0, 0, 0, 0,\ldots, 0, 0) \to (0, 1, 0, 1,\ldots, 0, 1) \to \\ & \qquad (1, 0, 1, 0,\ldots, 1, 0) \to (0, 0, 0, 0,\ldots, 0, 0).
\end{aligned}
\end{equation}

\begin{figure}[htp]
    \vspace{10pt}
    \centering
    \begin{tikzpicture}[
        scale=0.45,
        l/.style={draw=Black!30}, 
        s/.style={draw=Black, minimum size=12.6pt}, 
        e/.style={fill=White}, 
        o/.style={fill=Black}, 
        qg/.style={fill=QuasiparticleGreen}, 
        qr/.style={fill=QuasiparticleRed}, 
        qy/.style={fill=QuasiparticleYellow}, 
        ]
        \draw[l] (-0.5,-0.5) grid (7.5,8.5);
        \draw[l] (8.5,-0.5) grid (16.5,8.5);
        \draw[s, e] (0, 0) grid (7, 8);
        \node[s, o] at (0.5, 2.5) {};
        \node[s, o] at (0.5, 4.5) {};
        \node[s, o] at (0.5, 7.5) {};
        \node[s, o] at (1.5, 0.5) {};
        \node[s, o] at (1.5, 6.5) {};
        \node[s, o] at (2.5, 3.5) {};
        \node[s, o] at (3.5, 1.5) {};
        \node[s, o] at (3.5, 5.5) {};
        \node[s, o] at (3.5, 7.5) {};
        \node[s, o] at (4.5, 0.5) {};
        \node[s, o] at (4.5, 3.5) {};
        \node[s, o] at (4.5, 6.5) {};
        \node[s, o] at (5.5, 2.5) {};
        \node[s, o] at (5.5, 4.5) {};
        \node[s, o] at (6.5, 7.5) {};
        \draw[s, e] (9, 0) grid (16, 8);
        \node[s, o, fill=Black!20] at (9.5, 2.5) {};
        \node[s, o, fill=Black!20] at (9.5, 4.5) {};
        \node[s, o, fill=Black!20] at (9.5, 7.5) {};
        \node[s, o, fill=Black!20] at (10.5, 0.5) {};
        \node[s, o, fill=Black!20] at (10.5, 6.5) {};
        \node[s, o, fill=Black!20] at (11.5, 3.5) {};
        \node[s, o, fill=Black!20] at (12.5, 1.5) {};
        \node[s, o, fill=Black!20] at (12.5, 5.5) {};
        \node[s, o, fill=Black!20] at (12.5, 7.5) {};
        \node[s, o, fill=Black!20] at (13.5, 0.5) {};
        \node[s, o, fill=Black!20] at (13.5, 3.5) {};
        \node[s, o, fill=Black!20] at (13.5, 6.5) {};
        \node[s, o, fill=Black!20] at (14.5, 2.5) {};
        \node[s, o, fill=Black!20] at (14.5, 4.5) {};
        \node[s, o, fill=Black!20] at (15.5, 7.5) {};
        \node[s, qg] at (9.5, 0.5) {};
        \node[s, qg] at (9.5, 1.5) {};
        \node[s, qg] at (10.5, 1.5) {};
        \node[s, qg] at (10.5, 2.5) {};
        \node[s, qg] at (11.5, 1.5) {};
        \node[s, qg] at (11.5, 2.5) {};
        \node[s, qg] at (13.5, 4.5) {};
        \node[s, qg] at (13.5, 5.5) {};
        \node[s, qg] at (14.5, 5.5) {};
        \node[s, qg] at (14.5, 6.5) {};
        \node[s, qg] at (15.5, 5.5) {};
        \node[s, qg] at (15.5, 6.5) {};
        \node[s, qr] at (9.5, 5.5) {};
        \node[s, qr] at (9.5, 6.5) {};
        \node[s, qr] at (10.5, 4.5) {};
        \node[s, qr] at (10.5, 5.5) {};
        \node[s, qr] at (11.5, 4.5) {};
        \node[s, qr] at (11.5, 5.5) {};
        \node[s, qr] at (13.5, 1.5) {};
        \node[s, qr] at (13.5, 2.5) {};
        \node[s, qr] at (14.5, 0.5) {};
        \node[s, qr] at (14.5, 1.5) {};
        \node[s, qr] at (15.5, 0.5) {};
        \node[s, qr] at (15.5, 1.5) {};
        \node[s, qy] at (12.5, 2.5) {};
        \node[s, qy] at (12.5, 3.5) {};
        \node[s, qy] at (12.5, 4.5) {};
        \draw[->, >=stealth] (-1,2) -- node[left] {\large $i$} (-1,6);
        \draw[->, >=stealth] (1.5,-1) -- node[below] {\large $t$} (5.5,-1);
    \end{tikzpicture}
    \caption{\textbf{Interacting quasiparticles.} A fragment of a trajectory depicting the ballistic propagation and non-trivial interaction of quasiparticles. In the panel on the right, occupied sites are shaded. Green and red represent the location of the positive and negative solitons, respectively. That is, green and red coloured sites are those straddling domain walls between distinct vacua. The collision is coloured in yellow. Notice the transient speeding up of both solitons, which emerge from the collision further away from their original trajectories.  }
    \label{fig:quasiparticles}
\end{figure}
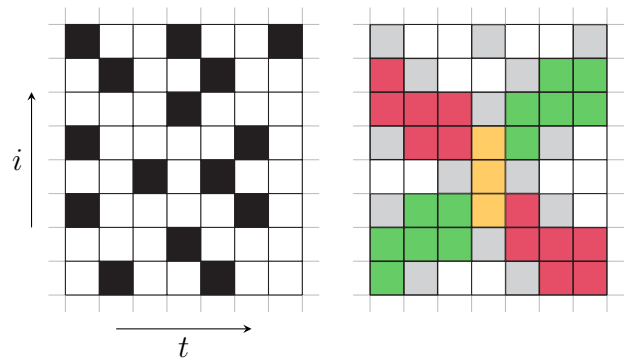

The quasiparticles, pairs of adjacent empty sites at the interfaces between vacua, propagate with an effective velocity of $\pm \frac{2}{3}$ and interact via a scattering process which effectively triples their velocity to $\pm 2$ for one time-step (see Fig.~\ref{fig:quasiparticles}). To distinguish the quasiparticles, we refer to them as either \textit{positive} or \textit{negative} depending on the sign of their velocity and denote their number within a configuration by the tuple,
\begin{equation}
    \q \equiv (\q^{+}, \q^{-}),
\end{equation}
where $\q^{\pm}$ denotes the number of positive and negative quasiparticles, respectively, in the configuration $\n$.

The quasiparticles can be detected diagrammatically by observing four consecutive sites of the lattice. If the binary string of these four adjacent sites reads either $(0, 0, 0, 1)$, $(1, 0, 0, 0)$, or $(1, 0, 0, 1)$ then a quasiparticle is present, as succinctly detailed by the following tables,
\begin{equation}\label{eq:quasiparticleTables}
\begin{array}{ c | c | c | c || c }
    \text{e} & \text{o} & \text{e} & \text{o} & \\
    \hline
    0 & 0 & 0 & 1 & - \\
    1 & 0 & 0 & 0 & - \\
    1 & 0 & 0 & 1 & + \\
\end{array}, \qquad
\begin{array}{ c | c | c | c || c }
    \text{o} & \text{e} & \text{o} & \text{e} & \\
    \hline
    0 & 0 & 0 & 1 & + \\
    1 & 0 & 0 & 0 & + \\
    1 & 0 & 0 & 1 & - \\
\end{array},
\end{equation}
where e/o denotes whether the adjacent sites indices are even or odd and $+/-$ whether the quasiparticle present is positive or negative. The quasiparticles can equivalently be identified by observing pairs of adjacent sites at the interfaces between vacua.

\begin{figure*}[ht!]
    \vspace{10pt}
    \centering
    \begin{tikzpicture}
        \node[thick, draw=Black, inner sep=0] at (0, 0) {\includegraphics[trim={282pt 22pt 282pt 24pt}, clip, width=.30\textwidth]{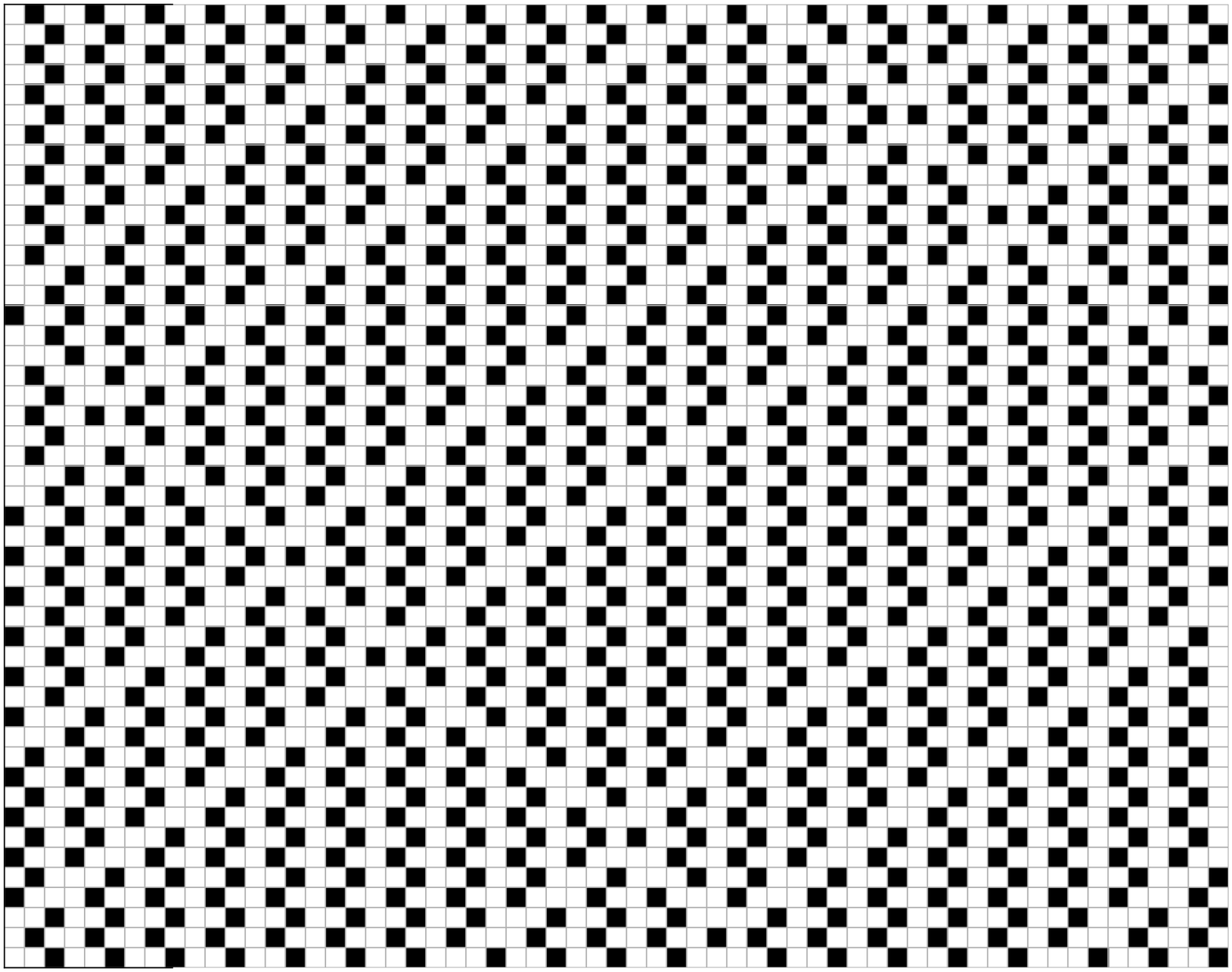}};
        \node[thick, draw=Black, inner sep=0] at (5.7, 0) {\includegraphics[trim={282pt 22pt 282pt 24pt}, clip, width=.30\textwidth]{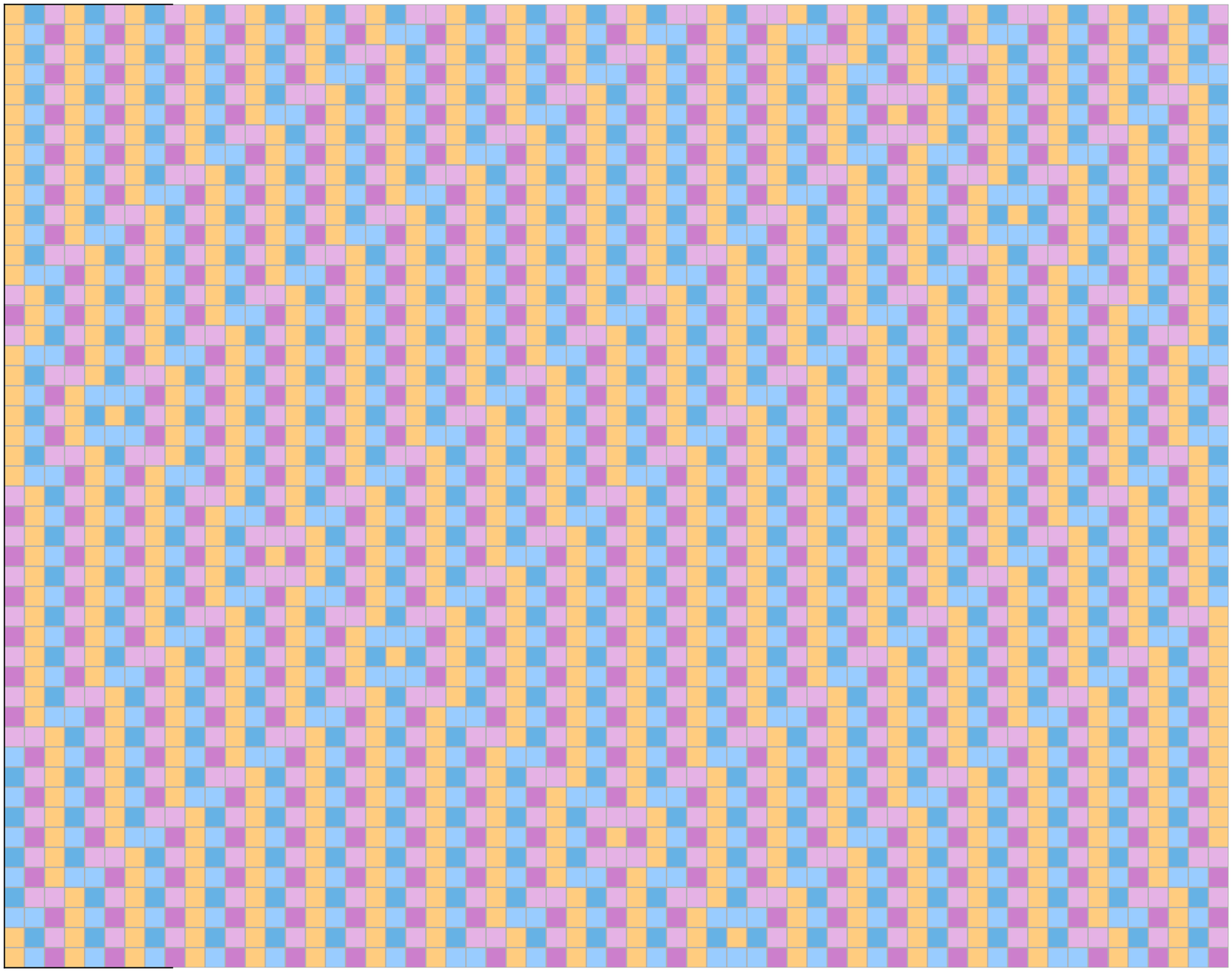}};
        \node[thick, draw=Black, inner sep=0] at (11.4, 0) {\includegraphics[trim={282pt 22pt 282pt 24pt}, clip, width=.30\textwidth]{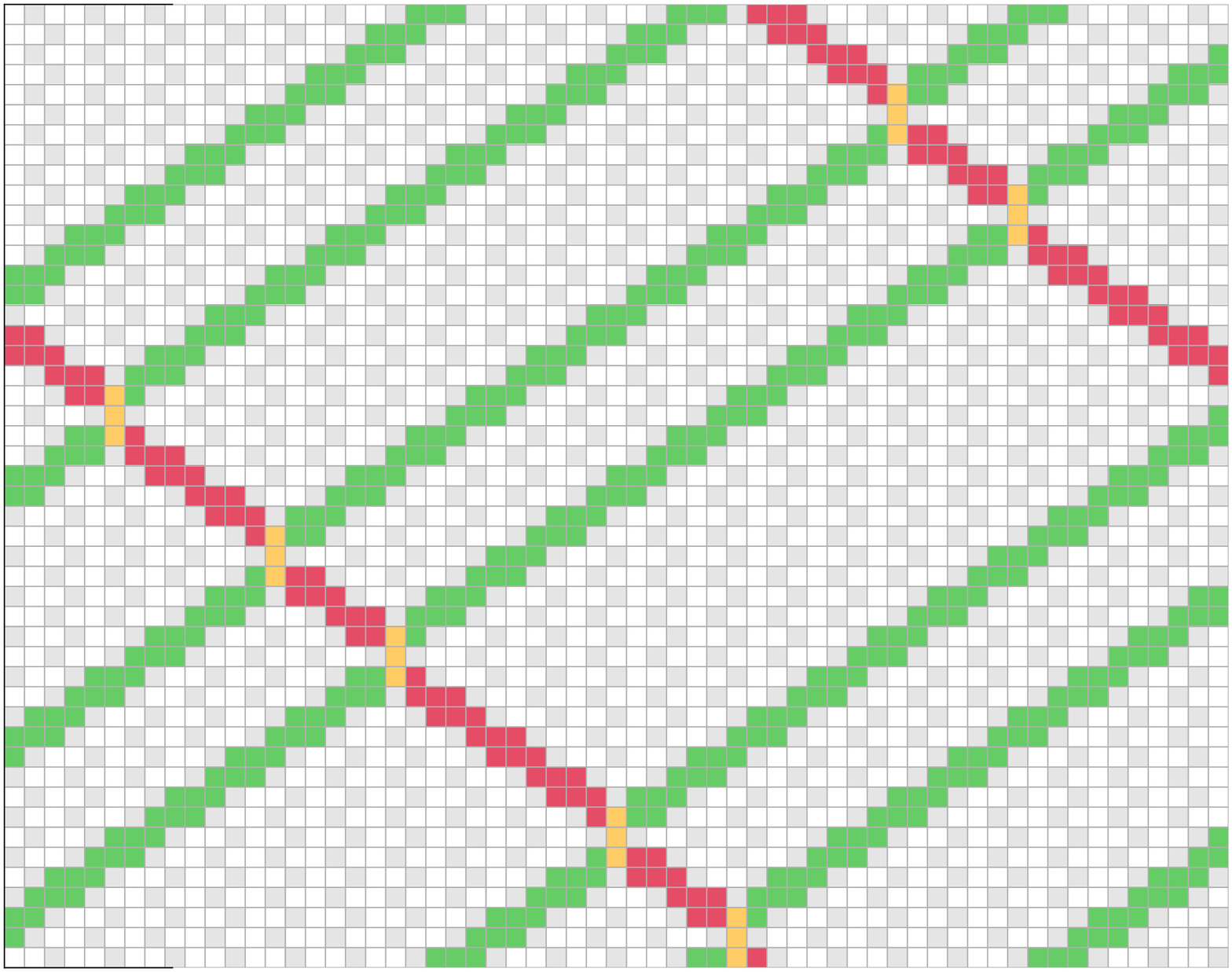}};
        \draw[->, >=stealth] (-2.89, -1) -- node[left] {\large $i$} (-2.89, 1);
        \draw[->, >=stealth] (-1, -2.32) -- node[below] {\large $t$} (1, -2.32);
    \end{tikzpicture}
    \caption{\textbf{Typical trajectory of the RCA201/Floquet-PXP.}
    A typical trajectory of the model in the subspace spanned by states with no adjacent occupied sites. The left panel represents the up and down sites as black and white, respectively. The middle panel shows the vacuum colour scheme, see Fig.~\ref{fig:vacua}. The right panel highlights the solitons. In this trajectory there are five solitons -- four positive movers and one negative -- that collide and wrap around the system due to the PBCs. Note that the location of the solitons coincides with domain walls between the vacuum states.}
    \label{fig:trajectory}
\end{figure*}

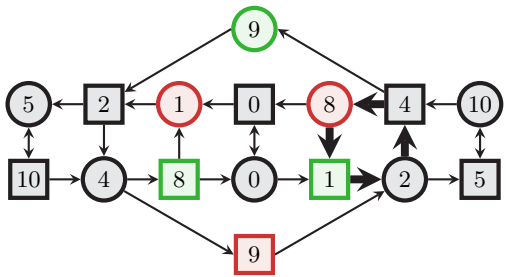
\begin{figure}[htp]
    \centering
    \vspace{10pt}
    \begin{tikzpicture}[
        scale=0.5,
        v/.style={ultra thick, inner sep=0pt, draw=Black, fill=Black!10, minimum size=14pt}, 
        e/.style={circle, minimum size=16pt}, 
        o/.style={rectangle}, 
        p/.style={draw=MyGreen, fill=MyGreen!10}, 
        n/.style={draw=MyRed, fill=MyRed!10}, 
        a/.style={thick, >=stealth, Black}, 
        ex/.style={ultra thick, >=stealth, MyOrange}, 
        ]
        \node[v, e] (e0) at (0,0) {$0$};
        \node[v, e, n] (e1) at (-2,2) {$1$};
        \node[v, e] (e2) at (4,0) {$2$};
        \node[v, e] (e4) at (-4,0) {$4$};
        \node[v, e] (e5) at (-6,2) {$5$};
        \node[v, e, n] (e8) at (2,2) {$8$};
        \node[v, e, p] (e9) at (0,4) {$9$};
        \node[v, e] (e10) at (6,2) {$10$};
        \node[v, o] (o0) at (0,2) {$0$};
        \node[v, o, p] (o1) at (2,0) {$1$};
        \node[v, o] (o2) at (-4,2) {$2$};
        \node[v, o] (o4) at (4,2) {$4$};
        \node[v, o] (o5) at (6,0) {$5$};
        \node[v, o, p] (o8) at (-2,0) {$8$};
        \node[v, o, n] (o9) at (0,-2) {$9$};
        \node[v, o] (o10) at (-6,0) {$10$};
        \draw[<->, a] (e0) -- (o0);
        \draw[->, a] (e0) -- (o1);
        \draw[->, a] (e1) -- (o2);
        \draw[->, a, line width=3pt] (e2) -- (o4);
        \draw[->, a] (e2) -- (o5);
        \draw[->, a] (e4) -- (o8);
        \draw[->, a] (e4) -- (o9.west);
        \draw[<->, a] (e5) -- (o10);
        \draw[->, a] (e8) -- (o0);
        \draw[->, a, line width=3pt] (e8) -- (o1);
        \draw[->, a] (e9.west) -- (o2);
        \draw[->, a] (e10) -- (o4);
        \draw[<->, a] (e10) -- (o5);
        \draw[->, a] (o0) -- (e1);
        \draw[->, a, line width=3pt] (o1) -- (e2);
        \draw[->, a] (o2) -- (e4);
        \draw[->, a] (o2) -- (e5);
        \draw[->, a, line width=3pt] (o4) -- (e8);
        \draw[->, a] (o4) -- (e9.east);
        \draw[->, a] (o8) -- (e0);
        \draw[->, a] (o8) -- (e1);
        \draw[->, a] (o9.east) -- (e2);
        \draw[->, a] (o10) -- (e4);
    \end{tikzpicture}
    \caption{\textbf{Quasiparticle number constraint.} Graph representation of the lattice illustrating the constraint \eqref{eq:quasiparticle-constraint} on the number of quasiparticles where, for readability, binary strings have been replaced by decimal integers (e.g. $(0,0,1,0) \equiv 2$). Vertices whose labels start on even and odd sites are represented by circles and squares with those denoting positive and negative quasiparticles in green and red. Black arrows then denote the directed edges between them. The cycle corresponding to an example configuration, $\n = (0,1,0,0)$, is indicated by bold arrows.}
    \label{fig:quasiparticle-constraint}
\end{figure}

Curiously, the numbers of positive and negative quasiparticles within any given configuration $\n$ are constrained and must satisfy the following equality,
\begin{equation}\label{eq:quasiparticle-constraint}
    \q^{+} - \q^{-} = 0 \pmod{3}.
\end{equation}
Naively, we can interpret this by postulating that the even system size $N$ and PBC impose that the quasiparticles exist as either positive-negative pairs or positive/negative triples. To prove this, we introduce a graph representation for the lattice, as illustrated in Fig.~\ref{fig:quasiparticle-constraint}. Specifically, we define a directed bipartite graph composed of two disjoint and independent sets of vertices, each identically labelled by binary strings of length four, and a set of directed edges between them. Here, the vertices of the two vertex sets represent the binary strings of consecutive sites within the lattice starting on even and odd sites, respectively, and the directed edges the possible transitions between them as the lattice is translated. We can simplify the graph by contracting paths along the directed edges between vertices whose binary labels denote quasiparticles. From here, with a relabelling of the vertices to denote positive and negative quasiparticles, it is trivial so see that any cycle of the graph satisfies Eq.~\eqref{eq:quasiparticle-constraint}.

\section{\label{sec:periodic} Exact stationary state for Periodic Boundary Conditions}
To study the macroscopic properties of the closed system we construct a class of macroscopic equilibrium states which we define as probability distributions over the set of configurations. For simplicity we will restrict most of the discussion to the configuration sector without pairs of up neighbours, in which case the numbers of both types of quasiparticles are conserved. (Being invariant, a cluster of two or more consecutive occupied sites acts as a reflective boundary for quasiparticles therefore changing their type but not their total number.) In this sector the simplest class of steady states can be constructed by introducing two chemical potentials, $\muP$ and $\muN$, associated with numbers of forward and backward moving quasiparticles respectively.

As we will demonstrate, such states can be expressed in two equivalent forms. We start by the \emph{patch state ansatz} (PSA) formulation of steady state, as introduced by~\cite{Prosen2016}. The main advantage of the PSA formulation is the construction, which can be done in absence of knowledge of conserved quantities, by simply requiring the states to be stationary and at the same time exhibit short-range correlations. Equivalently, the steady states can be expressed in terms of matrix product states (MPS). They obey a similar cubic algebraic relation to the MPS form of RCA54 steady states~\cite{Prosen2017}.

\subsection{\label{sec:master-equation} Macroscopic states and master equation}

We start the discussion of stationary states by first introducing the necessary formalism. Each configuration of the system~$\n$ is associated with the probability $p_{\n}$, that satisfies the non-negativity and normalization conditions,
\begin{equation}
  p_{\n}\ge0,\qquad \sum_{\{\n\}}p_{\n}=1.
\end{equation}
Each probability distribution, given by the set of probabilities $\{p_{\n}\}$, can be uniquely represented with a vector $\bm{p}\in(\mathbb{R}^2)^{\otimes N}$,
\begin{equation}
  \bm{p}=
  \sum_{\{\n\}} p_{\n} \bigotimes_{i=1}^{N} \bm{e}_{n_i},\qquad
  \bm{e}_{n} \equiv \begin{bmatrix}
    \delta_{n,0}\\
    \delta_{n,1}
  \end{bmatrix},
\end{equation}
where~$\bm{e}_0$ and $\bm{e}_1$ are the standard basis vectors of~$\mathbb{R}^2$. The state space is then identified as a convex subset of the vector space $(\mathbb{R}^{2})^{\otimes N}$.

The master equation describing the discrete time evolution of the system can be written as
\begin{equation}
  \bm{p}^{t+1}=\begin{cases}
    \Me \bm{p}^{t},& t = 0 \pmod{2},\\
    \Mo \bm{p}^{t},& t = 1 \pmod{2},
  \end{cases}
\end{equation}
where $\Me$ and $\Mo$ are transition matrices associated with the even and odd time-steps~\eqref{eq:even-odd-maps}, respectively,
\begin{equation}\begin{aligned}
  \bm{M}_{\text{O}} & :
    p_{n_1 n_2 \ldots n_{N - 1} n_{N}}
    \mapsto p_{f_1 n_2 \ldots f_{N - 1} n_{N}}, \\
    \bm{M}_{\text{E}} & :
    p_{n_{1} n_{2} \ldots n_{N - 1} n_{N}}
    \mapsto p_{n_{1} f_{2} \ldots n_{N - 1} f_N}.
\end{aligned}\end{equation}
The one time step propagators are equivalently given as products of local operators (gates),
\begin{equation}
  \begin{aligned}
    \Me & = \U_{2} \U_{4} \cdots \U_{N - 2} \U_{N}, \\
    \Mo & = \U_{1} \U_{3} \cdots \U_{N - 3} \U_{N - 1},
  \end{aligned}
\end{equation}
where for the bulk, $i \in \{2, \ldots, N - 1\}$,
\begin{equation}
  \U_{i} =
    \bm{I}^{\otimes (i - 2)} \otimes
    \U \otimes \bm{I}^{\otimes (N - i - 1)},
\end{equation}
are matrices encoding the deterministic bulk function in~\eqref{eq:local-update} (with the subscript indicating on which site of the lattice the operator acts non-trivially) whereas for the boundaries, $i \in \{1, N\}$, 
\begin{equation}\begin{aligned}
  \U_{1} & =
    \bm{I}^{\otimes N} + (\bm{X} - \bm{I}) \otimes
    \bm{P} \otimes \bm{I}^{\otimes (N - 3)} \otimes \bm{P}, \\
    \U_{N} & =
    \bm{I}^{\otimes N} + \bm{P} \otimes \bm{I}^{\otimes (N - 3)}
    \otimes \bm{P} \otimes (\bm{X} - \bm{I}),
\end{aligned}\end{equation}
are matrices encoding the left and right boundary functions, $f_{1}^{t}$ and $f_{N}^{t}$, respectively. Here,
\begin{equation}
  \U = \bm{I}^{\otimes 3} + \bm{P} \otimes (\bm{X} - \bm{I}) \otimes \bm{P},
\end{equation}
is the $8 \times 8$ permutation matrix enacting the local time evolution rule of Eq.~\eqref{eq:rule-201} on the vector space $(\mathbb{R}^{2})^{\otimes 3}$,
\begin{equation}
  \U =
    \begin{bmatrix}
        0 &   & 1 &   &   &   &   &   \\
          & 1 &   &   &   &   &   &   \\
        1 &   & 0 &   &   &   &   &   \\
          &   &   & 1 &   &   &   &   \\
          &   &   &   & 1 &   &   &   \\
          &   &   &   &   & 1 &   &   \\
          &   &   &   &   &   & 1 &   \\
          &   &   &   &   &   &   & 1 \\
    \end{bmatrix}.
\end{equation}
with $\bm{I}$, $\bm{P}$, and $\bm{X}$ the $2 \times 2$ identity, projector and Pauli-X matrices, respectively, acting on $\mathbb{R}^{2}$,
\begin{equation}
  \bm{I} =
    \begin{bmatrix}
      1 & 0 \\
        0 & 1 \\
    \end{bmatrix}, \qquad
    \bm{P} =
    \begin{bmatrix}
      1 & 0 \\
        0 & 0 \\
    \end{bmatrix}, \qquad
    \bm{X} =
    \begin{bmatrix}
      0 & 1 \\
        1 & 0 \\
    \end{bmatrix}.
\end{equation}

\subsection{Patch state ansatz formulation of Gibbs states}\label{sec:GibbsPSA}

We require a stationary state~$\bm{p}$ to map into itself after two time steps,
\begin{equation}
  \bm{p}=\Mo \Me \bm{p}.
\end{equation}
Due to $\U^{-1} = \U$, the stationarity condition can be equivalently recast as
\begin{equation}\label{eq:TIcond}
  \Mo \bm{p}=\Me \bm{p}.
\end{equation}
Similarly to the PSA introduced for RCA54 in~\cite{Prosen2016}, we propose the following form of the state~$\bm{p}$,
\begin{multline}
    p_{\n}\propto
    X^{\phantom{\prime}}_{n_1 n_2 n_3 n_4} X^{\prime}_{n_2 n_3 n_4 n_5}
    X^{\phantom{\prime}}_{n_3 n_4 n_5 n_6}\cdots\\
    \cdots 
    X^{\prime}_{n_{N-2} n_{N-1} n_N n_1}
    X^{\phantom{\prime}}_{n_{N-1} n_N n_1 n_2}
    X^{\prime}_{n_N n_1 n_2 n_3}.
\end{multline}
The values $X^{(\prime)}_{n_i n_{i+1} n_{i+2} n_{i+3}}$ are determined so that the stationarity condition in Eq.~\eqref{eq:TIcond} is satisfied. Explicitly, for any configuration~$\n=(n_1,n_2,n_3,\ldots,n_{N})$ the following equality has to hold,
\begin{multline}\label{eq:TIcondComp}
    X^{\phantom{\prime}}_{n_1 f_2 n_3 f_4}
    X^{\prime}_{f_2 n_3 f_4 n_5}\cdots
    X^{\prime}_{f_N n_1 f_2 n_3}\\
    = 
    X^{\phantom{\prime}}_{f_1 n_2 f_3 n_4}
    X^{\prime}_{n_2 f_3 n_4 f_5}\cdots
    X^{\prime}_{n_N f_1 n_2 f_3},
\end{multline}
where we have used the notation~$f_i=f(n_{i-1},n_i,n_{i+1})$, as introduced in~\eqref{eq:local-update}. Before solving the system of equations, we put all the components corresponding to configurations~$\n$ with pairs of consecutive $1$ to $0$ by requiring the following
\begin{equation}\label{eq:PSAsector}
  X^{(\prime)}_{11 n_1 n_2} = X^{(\prime)}_{n_1 1 1 n_2}
  = X^{(\prime)}_{n_1 n_2 1 1}=0.
\end{equation}
To fix the normalization we are free to choose $X_{0000}X^{\prime}_{0000}=1$, which together with~\eqref{eq:TIcondComp} implies
\begin{equation}\label{eq:PSAnchoice}
  X^{\phantom{\prime}}_{0101}X^{\prime}_{1010}=X^{\phantom{\prime}}_{1010}X^{\prime}_{0101}
  =X^{\phantom{\prime}}_{0000}X^{\prime}_{0000}=1.
\end{equation}
Additionally, we observe that the values $X^{(\prime)}_{n_1 n_2 n_3 n_4}$ are determined up to the following gauge transformation
\begin{equation}\label{eq:PSAgauge}
  \begin{aligned}
    X^{\phantom{\prime}}_{n_1 n_2 n_3 n_4} &\mapsto \alpha_{n_1 n_2 n_3}
    X^{\phantom{\prime}}_{n_1 n_2 n_3 n_4} \alpha^{\prime\,-1}_{n_2 n_3 n_4},\\
    X^{\prime}_{n_1 n_2 n_3 n_4} &\mapsto \alpha^{\prime}_{n_1 n_2 n_3}
    X^{\prime}_{n_1 n_2 n_3 n_4} \alpha^{-1}_{n_2 n_3 n_4},
  \end{aligned}
\end{equation}
which allows us to set $X^{(\prime)}_{0 n_1 n_2 n_3}=1$ for all configurations of three sites belonging to the sector without pairs of $1$,
\begin{equation}\label{eq:PSAgchoice}
  X^{(\prime)}_{0 n_1 n_2 n_3}=\left(1-\delta_{n_1+n_2,2}\right)\left(1-\delta_{n_2+n_3,2}\right).
\end{equation}
Combining the restriction to the relevant subspace~\eqref{eq:PSAsector} together with the choices of normalization~\eqref{eq:PSAnchoice} and gauge~\eqref{eq:PSAgchoice}, and requiring stationarity~\eqref{eq:TIcondComp} we obtain conditions for the remaining four components,
\begin{equation}
  X^{\phantom{\prime}}_{1000}=X^{\prime}_{1000}
  =X^{\phantom{\prime}}_{1001}X^{\prime}_{1001}.
\end{equation}
This condition exhibits the following two-parameter family of solutions,
\begin{equation}
  \begin{gathered}
    X^{\phantom{\prime}}_{1001}=\frac{\omega^2}{\xi},\qquad
    X^{\prime}_{1001}=\frac{\xi^2}{\omega},\\
    X^{\phantom{\prime}}_{1000}=X^{\prime}_{1000} =\omega\xi,
  \end{gathered}
\end{equation}
with all the other components either being $0$ (as given by~\eqref{eq:PSAsector}) or~$1$. The vector~$\bm{p}$ representing the steady state has to be normalized, therefore all its components have to be non-negative, which restricts the values of parameters~$\xi$, $\omega$ to~$\mathbb{R}^{+}$.

At this point the choice of parametrization is arbitrary, but it can be straightforwardly demonstrated that the parameters $\xi$ and $\omega$ are exponents of the chemical potentials $\muP$, $\muN$ corresponding to numbers of positively and negatively moving quasiparticles. First we use the gauge freedom to transform the tensors into an equivalent form,
\begin{equation}
  \begin{gathered}
    \alpha_{000}=1,\qquad
    \alpha_{010}=\xi^{-1},\\
    \alpha_{001}=\alpha_{100}=\alpha_{101}=\omega^{-1},\\
    \alpha^{\prime}_{n_1 n_2 n_3}=\left.\alpha_{n_1 n_2 n_3}\right|_{\xi\leftrightarrow \omega}, \\
  \end{gathered}
\end{equation}
which by~\eqref{eq:PSAgauge} implies
\begin{equation}
  \begin{aligned}
    X^{\phantom{\prime}}_{0001}&\mapsto\xi,\\
    X^{\phantom{\prime}}_{1000}&\mapsto\xi,\\
    X^{\phantom{\prime}}_{1001}&\mapsto\omega,
  \end{aligned}\qquad
  \begin{aligned}
    X^{\prime}_{0001}&\mapsto\omega,\\
    X^{\prime}_{1000}&\mapsto\omega,\\
    X^{\prime}_{1001}&\mapsto\xi,
  \end{aligned}
\end{equation}
while the other components either remain $0$, cf.~\eqref{eq:PSAsector}, or are mapped into~$1$. In a given configuration~$\n$, the number of both types of quasiparticles can be determined by the count of sub-configurations $(0,0,0,1)$, $(1,0,0,0)$ and $(1,0,0,1)$. Depending on the parity of the site indices where the sub-configurations are positioned, they correspond either to quasiparticles with positive or negative velocity, as summarized by Eq.~\eqref{eq:quasiparticleTables}. Therefore, the new values of~$X^{(\prime)}_{n_1 n_2 n_3 n_4}$ imply that every component~$p_{\n}$ of the stationary state~$\bm{p}$ is weighed as
\begin{equation}
  p_{\n} \propto
  \xi^{\q^{+}} \omega^{\q^{-}},
\end{equation}
where~$\q^{\pm}$ are the numbers of positive and negative quasiparticles in a given configuration $\n$.

Since the requirement for stationarity is the invariance to evolution for two time-steps~\eqref{eq:TIcond}, we can define two versions of state,~$\bm{p}$ and~$\bm{p}^{\prime}$, corresponding to even and odd time-steps respectively,
\begin{equation}\label{eq:TIcondTwoStates}
  \bm{p}^{\prime}=\Me\bm{p},\qquad \bm{p}=\Mo\bm{p}^{\prime}.
\end{equation}
Together with the solution for~$\bm{p}$, this condition implies that the odd-time version of the state takes the same form with the roles of~$X^{\phantom{\prime}}_{n_1 n_2 n_3 n_4}$ and $X^{\prime}_{n_1 n_2 n_3 n_4}$ reversed,
\begin{multline}
  p^{\prime}_{\n}\propto
  X^{\prime}_{n_1 n_2 n_3 n_4} X^{\phantom{\prime}}_{n_2 n_3 n_4 n_5}
  X^{\prime}_{n_3 n_4 n_5 n_6}\cdots\\
  \cdots
  X^{\phantom{\prime}}_{n_{N-2} n_{N-1} n_N n_1}
  X^{\prime}_{n_{N-1} n_N n_1 n_2}
  X^{\phantom{\prime}}_{n_N n_1 n_2 n_3}.
\end{multline}
This parametrization of the steady state preserves the symmetry of the model: shifting the state by one site (up or down) is the same as evolving it for one time-step (half of Floquet period).

\subsection{Matrix product form of stationary states}

Equivalently, the stationary states can be recast in the matrix product form,
\begin{equation}\label{eq:MPSdefinition}
  \bm{p}=\frac{1}{Z}\tr\Big(
    \vW_1 \vV_2 \vW_3
  \cdots \vW_{N-1} \vV_N\Big),
\end{equation}
where~$\vWV_i$ are vectors of matrices, corresponding to the physical site $i$, $\vWV=\big(\WV_0,\WV_1\big)^T$, and $Z$ is the normalization. Explicitly, the components~$p_{\n}$ of the state~$\bm{p}$ read
\begin{equation}
  p_{\n}=\frac{1}{Z}\tr\Big( \W_{n_1} \V_{n_2} \W_{n_3}\cdots
    \W_{n_{N-1}}
  \V_{n_N}\Big).
\end{equation}
To construct MPS from the PSA, we introduce an $8$-dimensional auxiliary space with each basis element labeled by a binary string $(m_1 m_2 m_3)$ and we define the $8\times 8$ matrices $\tWV_n$ with the entries given by the PSA values as
\begin{equation}
  (\tWV_{n})^{m_1^{\prime} m_2^{\prime}m_3^{\prime}}_{m_1 m_2 m_3}
    = \delta_{m_1^{\prime},m_2}\delta_{m_2^{\prime},m_3}\delta_{m_3^{\prime},n}
    X^{(\prime)}_{m_1 m_2 m_3 n},
\end{equation}
where the strings in the superscript and the subscript are the binary representations of the row and column index respectively. MPS consisting of these matrices are equivalent to the PSA steady state as introduced before,
\begin{equation}\label{eq:MPStilde}
  \begin{aligned}
    \tr\Big(
      \tW_{n_1}
      \tV_{n_2}\cdots
      \tV_{n_N}
    \Big)=
    X_{n_1 n_2 n_3 n_4}
    \cdots
    X^{\prime}_{n_N n_1 n_2 n_3}.
  \end{aligned}
\end{equation}
The MPS can be simplified by introducing $4\times 8$ and $8\times4$ auxiliary space matrices~$R$ and $Q$
\begin{equation}
  R=\begin{bmatrix}
    1& & & &0& & & \\
    &1& & & &1& & \\
    & &1& & & &0& \\
    & & &0&1& & &0
  \end{bmatrix},\quad
  Q=\begin{bmatrix}
    1& & & \\
    &1& & \\
    & &1& \\
    & & &0\\
    0& & &1\\
    &0& & \\
    & &0& \\
    & & &0
  \end{bmatrix},
\end{equation}
and noting that for any combination of $n_1$, $n_2$, inserting $QR$ between two consecutive matrices does not change the product,
\begin{equation}
  \tW_{n_1} Q R \tV_{n_2}
  = \tW_{n_1} \tV_{n_2}.
\end{equation}
From here it follows that the MPS~\eqref{eq:MPSdefinition} composed of $4\times 4$ matrices~$\WV_n$, defined as~$\WV_n=R\tWV_n Q$, is equivalent to~\eqref{eq:MPStilde}. Explicitly,
\begin{equation}
  \W_0=\begin{bmatrix}
    1&0&0&\xi\\
    0&0&0&0\\
    0&1&0&0\\
  0&0&1&0 \end{bmatrix},\qquad
  \W_1=\begin{bmatrix}
    0&0&0&0\\
    \xi&0&1&\omega\\
    0&0&0&0\\
    0&0&0&0
  \end{bmatrix},
\end{equation}
while the other pair of matrices is given by the exchange of parameters $\xi\leftrightarrow\omega$,
\begin{equation}
  \V_n(\xi,\omega)=\W_n(\omega,\xi).
\end{equation}

The stationarity of the MPS is implied by the equivalence between the two representations. However, the MPS additionally exhibits an algebraic structure that allows us to explicitly demonstrate the stationarity without relying on the equivalence with the PSA. Matrices $\WV_n$ satisfy a cubic algebraic relation, analogous to~\cite{Prosen2017},
\begin{equation}\label{eq:3siteAlgebra}
  \U_{2}
  \Big(\vW_1 \vV_2 \vW_3 S\Big)
  =\vW_1 S\, \vW_2 \vV_3,
\end{equation}
which compactly encodes the following component-wise equalities,
\begin{equation}
  \W_{n_1}\V_{f(n_1,n_2,n_3)}\W_{n_3} S
  = \W_{n_1} S\, \W_{n_2}\V_{n_3}.
\end{equation}
We introduced the \emph{delimiter matrix} $S$, defined as
\begin{equation}
  S=\begin{bmatrix}
    \frac{\xi \omega}{\xi^2-\omega} & -\frac{\omega}{\xi^2-\omega}
    & 0 & \frac{\xi^2}{\xi^2-\omega}\\
    1&0&0&\omega\\
    0&0&1&0\\
    -\frac{\omega}{\xi^2-\omega}& \frac{\xi}{\xi^2-\omega}&0&
    -\frac{\xi}{\xi^2-\omega}
  \end{bmatrix}.
\end{equation}
The inverse of the delimiter matrix is given by exchanging the parameters,
\begin{equation}
  S(\xi,\omega)^{-1}=S(\omega,\xi),
\end{equation}
which immediately implies a \emph{dual} relation similar to~\eqref{eq:3siteAlgebra},
\begin{equation}\label{eq:3siteDualAlgebra}
  \U_{2}\Big(\vV_1\vW_2\vV_3 S^{-1}\Big)
  = \vV_1 S^{-1} \vV_2\vW_3.
\end{equation}
Note that in case $\xi=\omega^2$ or $\omega=\xi^2$, the matrices~$S$ and $S^{-1}$ are not well defined, however the products~$V^{\phantom{\prime}}_{n}S$ and $V^{\prime}_n S^{-1}$ have finite values in the limit~$\xi\to\omega^2$ (or $\omega\to\xi^2$). Therefore the following discussion holds for any value of parameters. When $\xi=\omega=1$, the stationary state becomes the maximum entropy state, where each allowed configuration is equally likely. In this case the MPS representation can be reduced to~$2\times 2$ matrices, as is explained in Appendix~\ref{sec:MEstate}.

The odd-time version of the state, $\bm{p}^{\prime}$, has the same form as~$\bm{p}$, but the parameters~$\xi$ and $\omega$ are exchanged (or equivalently, $\vV$ is replaced by $\vW$ and vice versa),
\begin{equation} \label{eq:GibbsPrime}
  \bm{p}^{\prime}
    =\frac{1}{Z}\tr\Big( \vV_1 \vW_2 \vV_3 \cdots
      \vV_{N-1} \vW_N \Big).
\end{equation}
The stationarity requirement~\eqref{eq:TIcondTwoStates} follows directly from  relations~\eqref{eq:3siteAlgebra} and \eqref{eq:3siteDualAlgebra}. To prove the first of the stationarity conditions, we insert $S S^{-1}$ between the matrices corresponding to the first and second sites, and apply the local time evolution operator~$\U_{N}$ using the $3$-site algebraic relation,
\begin{equation}
  \begin{aligned}
    \Me
    &\tr\Big( \vW_1 \vV_2 \vW_3
    \cdots \vW_{N-1} \vV_N \Big)\\
    =
    \smashoperator{\prod_{i=1}^{N/2}}
    \U_{2i}
    &\tr\Big( \vW_{N-1} \vV_{N} \vW_1
    S S^{-1} \vV_2 \cdots \vW_{N-3}\vV_{N-2} \Big)\\
    =\smashoperator{\prod_{i=1}^{N/2-1}}
    \U_{2i}
    &\tr\Big( \vV_1 S^{-1} \vV_2 \cdots
    \vW_{N-3} \vV_{N-2} \vW_{N-1} S \vW_{N} \Big).
  \end{aligned}
\end{equation}
We keep applying local time evolution operators $\U_{N-2}$, $\U_{N-4}$, \ldots, one by
one, each time moving the matrix~$S$ two sites to the left as described by~\eqref{eq:3siteAlgebra}, until we are left with the following
\begin{equation}
  \begin{aligned}
    \U_{2}&\tr\Big( \vV_1 S^{-1}\vV_2 \vW_3 S
      \vW_4\cdots \vV_{N-1} \vW_{N} \Big)\\
      = &\tr\Big(\vV_1 \vW_2 \vV_3 S^{-1}S \vW_4\cdots
      \vV_{N-1} \vW_{N} \Big),
  \end{aligned}
\end{equation}
where we used the dual relation in Eq.~\eqref{eq:3siteDualAlgebra} together with~$\U^{-1}=\U$. Thus we proved that the even time evolution operator~$\Me$ maps the state~$\bm{p}$ into its odd-time analogue~$\bm{p}^{\prime}$. The second stationarity requirement~\eqref{eq:TIcondTwoStates} can be proved analogously.

\subsection{\label{sec:partition-function-quasiparticles} Partition function}

As demonstrated in Subsec.~\ref{sec:GibbsPSA}, the stationary probabilities of configurations~$p_{\n}$ are distributed according to the grand-canonical ensemble,
\begin{equation}\label{eq:quasiparticle-gibbs-state}
    p_{\n} = \frac{1}{Z} \exp \big( \q^{+} \mu^{+} + \q^{-} \mu^{-} \big),
\end{equation}
with the chemical potentials corresponding to the numbers of positive and negative quasiparticles determined by the parameters
\begin{equation}
    \xi = \e^{\mu^{+}}, \qquad \omega = \e^{\mu^{-}}.
\end{equation}
The partition function $Z$ can therefore be given in two equivalent forms. The first one follows directly from the normalization condition of the MPS representation of the stationary state~$\bm{p}$
\begin{equation}\label{eq:partition-function2}
    Z=\sum_{\{\n\}} \tr \big(\W_{n_1} \V_{n_2} \W_{n_3}\cdots \V_{n_N}\big)\equiv\tr T^{N/2},
\end{equation}
where we introduced the transfer matrix $T$ as the sum of all products of matrices on two sites,
\begin{equation}\label{eq:transfer-matrix-def}
    T = (V^{}_{0} + V^{}_{1})(V^{\prime}_{0} + V^{\prime}_{1}) =
    \begin{bmatrix}
        1 & 0 & \xi & \omega \\
        \xi & 1 & \omega & \xi \omega \\
        \omega & 0 & 1 & \xi \\
        0 & 1 & 0 & 0 \\
    \end{bmatrix}.
\end{equation}
The second form of $Z$ is defined as a weighted sum over the set of quasiparticle numbers,
\begin{equation}\label{eq:partition-function}
    Z = \sum_{\{\n\}} \xi^{\q^{+}} \omega^{\q^{-}} = \sum_{\{Q\}} \Omega_{Q} \xi^{Q^{+}} \omega^{Q^{-}},
\end{equation}
where the entropic term $\Omega_{Q} = \Omega(N, Q^{+}, Q^{-})$, which counts the number of degenerate configurations with the same number of quasiparticles, takes the following combinatoric form
\begin{equation}\label{eq:quasiparticle-entropy}
    \Omega_{Q} = \frac{1}{m_{Q}} \binom{\frac{1}{2} N - \frac{1}{3} Q^{+} - \frac{2}{3} Q^{-}}{Q^{+}} \binom{\frac{1}{2} N - \frac{1}{3} Q^{-} - \frac{2}{3} Q^{+}}{Q^{-}},
\end{equation}
with $m_{Q} = m(N, Q^{+}, Q^{-})$ the \textit{time-averaged magnetization density} expressed in terms of the numbers of positive and negative quasiparticles as
\begin{equation}
    m_{Q} = \frac{\big(\frac{1}{2} N - \frac{1}{3} Q^{+} - \frac{2}{3} Q^{-}\big)\big(\frac{1}{2} N - \frac{1}{3} Q^{-} - \frac{2}{3} Q^{+}\big)}{\frac{3}{2} N \big(\frac{1}{2} N - \frac{2}{3} Q^{+} - \frac{2}{3} Q^{-}\big)}.
\end{equation}
The set $\{Q\}$ above denotes the set of tuples of numbers of positive and negative quasiparticles that satisfy both the equality in Eq.~\eqref{eq:quasiparticle-constraint} imposed by the even system size and PBC and the following inequalities that manifest from the finite effective size of the quasiparticles,
\begin{equation}\label{eq:quasiparticle-size-constraint}
    Q^{\pm} + 2 Q^{\mp} \leq \frac{3}{2} N,
\end{equation}
which is implicitly given by $\binom{n<k}{k}=0$. To prove that the expression~\eqref{eq:quasiparticle-entropy} really represents the entropic contribution, it suffices to show that the two forms of the partition sum (given by Eqs.~\eqref{eq:partition-function2} and~\eqref{eq:partition-function}) coincide. The proof of equivalence is provided in Appendix~\ref{sec:partition-function-equivalence}. 

Alternatively, the inequalities of Eq.~\eqref{eq:quasiparticle-size-constraint} can be understood directly from the quasiparticle picture. First we consider the minimum effective size of the pairs and triples of quasiparticles (i.e. the minimum number of sites they occupy within a configuration). Noting from inspection that they cover at least four and eight sites, respectively, we obtain the following expression,
\begin{equation}
\label{eq:Qcons}
    4 Q^{(2)} + 8 Q^{(3)} \leq N,
\end{equation}
where $Q^{(2)}$ and $Q^{(3)}$ denote the numbers of pairs and triples of quasiparticles, respectively. We now express these in terms of the numbers of positive and negative quasiparticles, where for $Q^{\pm} \geq Q^{\mp}$, we have
\begin{equation}
\label{eq:Qcons2}
    Q^{(2)} = Q^{\mp}, \qquad Q^{(3)} = \frac{1}{3}\big(Q^{\pm} - Q^{\mp}\big).
\end{equation}
A simple substitution then yields the inequalities outlined in Eq.~\eqref{eq:quasiparticle-size-constraint}.

In the limit of large $N$ the expression for the partition function \eqref{eq:partition-function} can be written in terms of an integral over {\em quasiparticle densities},
\begin{equation}
    \rho^{\pm} = \frac{Q^{\pm}}{N},
\end{equation}
to read
\begin{equation}
    Z = \int_{0}^1 d\rho^{+} d\rho^{-} \exp \left(N {\cal F}(\rho^{+},\rho^{-})\right),
\end{equation}
where ${\cal F}$ is (minus) a free energy density with ``energetic'' terms, associated with the cost of each soliton species in terms of their chemical potential, and entropic terms from the counting of states,
\begin{equation}
    {\cal F} = \mu^{+} \rho^{+} + \mu^{-} \rho^{-} + {\cal S}(\rho^{+},\rho^{-}) .  
\end{equation}
The entropy density ${\cal S}$ is obtained from using the Stirling approximation in \eqref{eq:quasiparticle-entropy}. It reads
\begin{equation}
\begin{aligned}
    \mathcal{S} = & -\rho^{+} \ln \rho^{+} \\ 
    & + \bigg(\frac{1}{2} - \frac{1}{3} \rho^{+} - \frac{2}{3} \rho^{-}\bigg) \ln \bigg(\frac{1}{2} - \frac{1}{3} \rho^{+} - \frac{2}{3} \rho^{-}\bigg) \\ 
    & - \bigg(\frac{1}{2} - \frac{2}{3} \rho^{-} - \frac{4}{3} \rho^{+}\bigg) \ln \bigg(\frac{1}{2} - \frac{2}{3} \rho^{-} - \frac{4}{3} \rho^{+}\bigg) \\
    & + \left( \rho^{+} \leftrightarrow \rho^{-} \right),
\end{aligned}
\end{equation}
and has the form of an entropy density of mixing of the quasiparticles subject to the constraints \eqref{eq:Qcons} and \eqref{eq:Qcons2}.

\section{\label{sec:stochastic} Exact stationary state for Stochastic Boundary Conditions}
The RCA201/Floquet-PXP with PBC is fully deterministic. The integrability of the model implies that the dynamics is naturally decomposed into many different sectors, which makes the number of steady states of the closed system highly degenerate. In the absence of chaos, a way to make the dynamics ergodic is to impose stochastic boundary conditions (SBCs) by considering a finite chain coupled to stochastic reservoirs on both ends, an approach similar to that of the RCA54, cf.~\cite{Prosen2016,Inoue2018,Prosen2017}. With SBCs the RCA201/Floquet-PXP becomes a stochastic model, and by ergodic we mean two things. First, all configurations are dynamically connected, that is, the relevant subspace is irreducible under the dynamics since quasiparticles can be created and destroyed at the boundaries. Note that this subspace is slightly larger than that of a similarly sized system with PBCs as with SBCs there is no restriction on the occupation of the first and last site which are no longer neighbours. The number of configurations in the subspace of interest is then the Fibonacci rather than the Lucas number (see Subsec.~\ref{sec:space}). Second, the relaxation time (i.e. the time to forget a typical initial condition) is finite. 

In this section we find a class of suitable stochastic boundary propagators to make the system relax to a unique \emph{non-equilibrium steady state} (NESS) similar to the Gibbs state introduced in Sec.~\ref{sec:periodic}. The starting point is the MPS form of the Gibbs state of a large system with periodic boundaries, which is used to express the probability distribution (i.e.\ state) of a finite subsection of the chain in the limit when the system size goes to infinity. The resulting probability distribution can be viewed as a NESS of the finite chain with the boundaries that stochastically inject and remove quasiparticles with rates that are compatible with the chemical potentials~$\muP$, $\muN$ of the original Gibbs state.

\subsection{State of a finite section of a larger system}\label{subsec:asymptDistr}

We start with the closed system with periodic boundary conditions and length~$M$ that is assumed to be the equilibrium state given by spectral parameters~$\xi$, $\omega$, as introduced in Sec.~\ref{sec:periodic}. By definition, the probabilities of configurations of a smaller section of the chain with length~$N$ are given by summing over the probabilities corresponding to the configurations $(n_1,n_2\ldots n_{M})$ with the same first~$N$ bits,
\begin{equation}
  p^{(M)}_{n_1 \ldots n_N}=
  \smashoperator{\sum_{n_{N+1}\ldots n_M}}
  Z^{-1}\tr\Big(\W_{n_1} \V_{n_2} \cdots
  \V_{n_M}\Big)
\end{equation}
Note that the superscript~$(M)$ refers to the length of the whole system and not the length of the section. Using~$T$ to denote the transfer matrix, $T=(\W_0+\W_1)(\V_0+\V_1)$, as introduced in Eq.~\eqref{eq:transfer-matrix-def}, the probability distribution~$\bm{p}^{(M)}$ can be succinctly expressed as
\begin{equation}
  \begin{aligned}
    \bm{p}^{(M)}
    =\frac{\tr\Big(\vW_{1} \vV_{2} \cdots
    \vV_{N} T^{(M-N)/2}\Big)}{\tr T^{M/2}}.
  \end{aligned}
\end{equation}
We define the state of the subsystem~$\bm{p}$ as the large system size limit of the distribution~$\bm{p}^{(M)}$,
\begin{equation}\label{eq:asymptNess}
  \bm{p}=\lim_{M\to\infty} \bm{p}^{(M)}
  =\frac{
    \bra{l}\vW_1\vV_2\cdots\vV_{N}\ket{r}
  }{\lambda^{N/2}\braket{l}{r}},
\end{equation}
where we introduced the parameter~$\lambda$ denoting the leading eigenvalue of the matrix~$T$, and $\bra{l}$, $\ket{r}$ are the corresponding left and right eigenvectors,
\begin{equation}
  T\ket{r}=\lambda\ket{r},\qquad \bra{l}T=\lambda \bra{l}.
\end{equation}
Explicitly, $\lambda$ is the largest solution of the following quartic equation,
\begin{multline}
  \lambda^4-3\lambda^3+(3-2\xi\omega)\lambda^2-(1-\xi\omega)\lambda
  \\-(\xi^2-\omega)(\omega^2-\xi)=0,
\end{multline}
while the leading eigenvectors are implicitly given by parameters $\xi$, $\omega$ and the eigenvalue $\lambda$ as
\begin{equation}
  \bra{l}=((\lambda-1)\xi+\omega^2)
  \begin{bmatrix}
    (\lambda-1)\xi+\omega^2\\
    (\lambda-1)^2-\xi \omega\\
    (\lambda-1)\omega+\xi^2\\
    (\lambda-1)\left((\lambda-1)^2-\xi\omega\right)
  \end{bmatrix}^T,
\end{equation}
and
\begin{equation}
  \ket{r}=
  ((\lambda-1)\omega+\xi^2)
  \begin{bmatrix}
    \lambda\left( (\lambda-1)^2-\xi\omega\right)\\
    \lambda\left( (\lambda-1)\xi+\omega^2\right)\\
    \lambda(\lambda-1)\omega-\xi\omega^2+\xi^2\\
    (\lambda-1)\xi+\omega^2
  \end{bmatrix},
\end{equation}
where the nontrivial normalization prefactor is chosen to simplify the boundary equations in the next subsection. Note that the asymptotic form of the probability distribution~\eqref{eq:asymptNess} is valid as long as the leading eigenvalue $\lambda$ is not degenerate, which is the case for all $\xi,\omega > 0$. The odd time-step version of the asymptotic distribution, $\bm{p}^{\prime}$, takes the same form as~$\bm{p}$ with the exchanged roles of parameters $\xi$ and $\omega$. Explicitly,
\begin{equation}\label{eq:pAsymptNess}
\bm{p}^{\prime}
  =\frac{
    \bra{l^{\prime}}\vV_1\vW_2\cdots
    \vW_{N}\ket{r^{\prime}}
  }{\lambda^{N/2}\braket{l^{\prime}}{r^{\prime}}},
\end{equation}
where the vectors~$\bra{l^{\prime}}$ and $\ket{r^{\prime}}$ are defined as
\begin{equation}
  \bra{l^{\prime}(\xi,\omega)}=\bra{l(\omega,\xi)},\qquad
  \ket{r^{\prime}(\xi,\omega)}=\ket{r(\omega,\xi)},
\end{equation}
and the leading eigenvalue~$\lambda$ is invariant under the exchange $\xi\leftrightarrow\omega$.

To avoid the cluttering of notation, we use the symbols~$\bm{p}$, $\bm{p}^{\prime}$ to denote probability distributions on~$N$ sites, i.e.\ $\bm{p}^{(\prime)}$ are vectors from~$(\mathbb{R}^{2})^{\otimes N}$ with components~$p^{(\prime)}_{n_1 n_2 n_3\ldots n_N}$. When we refer to probabilities of configurations of different lengths, we will always use the component-wise notation to avoid ambiguity. Note that values $p^{(\prime)}_{n_1 n_2\ldots n_k}$ take the form similar to~\eqref{eq:asymptNess} and~\eqref{eq:pAsymptNess} with $N$ being replaced by~$k$.

\subsection{Compatible boundaries}

The probability distribution of the section of the chain, $\bm{p}$, can be understood as the NESS of a boundary driven system. We assume the one time-step evolution operators to be deterministic in the bulk and stochastic at the boundaries. Explicitly, under the even time-step operator~$\Me$ the sites $(1,2,\ldots,N-4)$ change deterministically according to the time evolution rule~\eqref{eq:rule-201}, while the evolution of sites~$(N-3,N-2,N-1,N)$ is given by a stochastic matrix~$\RR$,
\begin{equation}
    \Me=\smashoperator{\prod_{i=1}^{N/2-2}}\U_{2i}
    \,\RR_{N-3 N-2 N-1 N}.
\end{equation}
Similarly, in the odd time-step, the evolution of sites $(5,6,7,\ldots,N)$ is deterministic and the evolution of the first four sites $(1,2,3,4)$ is encoded in the stochastic matrix~$\LL$,
\begin{equation}
    \Mo=\LL_{1234}\smashoperator{\prod_{i=2}^{N/2-1}}\U_{2i+1}.
\end{equation}
For the vectors~$\bm{p}$, $\bm{p}^{\prime}$ to be understood as a stationary state under the stochastic time evolution, the following conditions have to be satisfied,
\begin{equation}
  \Me\bm{p}=\bm{p}^{\prime},\qquad \Mo\bm{p}^{\prime}=\bm{p}.
\end{equation}
The stationarity condition is fulfilled when in addition to the bulk algebraic relations~\eqref{eq:3siteAlgebra}, the MPS introduced in~\eqref{eq:asymptNess} and~\eqref{eq:pAsymptNess} satisfies the appropriate boundary relations. Explicitly,~$\bm{p}$ is mapped into~$\bm{p}^{\prime}$ under the even time-step evolution, when the following boundary equations hold,
\begin{equation}\label{eq:boundaryAlgebra1}
  \begin{aligned}
    \bra{l}\vW_1 S
    &= \Gamma \bra{l^{\prime}} \vV_1, \\
    \RR_{1234}\Big(\vW_1 \vV_2 \vW_3 \vV_4\ket{r}\Big)
    &=\vW_1 S \vW_2 \vV_3 \vW_4\ket{r^{\prime}}.
  \end{aligned}
\end{equation}
Analogously, the second stationarity condition implies the following two boundary relations,
\begin{equation}\label{eq:boundaryAlgebra2}
  \begin{aligned}
    \LL_{1234}\Big(\bra{l^{\prime}}\vV_1 \vW_2 \vV_3 \vW_4 \Big)
    &= \bra{l}\vW_1\vV_2\vW_3\vV_4 S^{-1},\\
    \vV_1 S^{-1}\ket{r^{\prime}} &= \frac{1}{\Gamma} \vV_1\ket{r},
  \end{aligned}
\end{equation}
where the scalar factor~$\Gamma$ is determined by the normalisation of the MPS as
\begin{equation}
  \Gamma=\frac{\braket{l}{r}}{\braket{l^{\prime}}{r^{\prime}}}
  =\frac{(\lambda-1)\xi+\omega^2}{(\lambda-1)\omega+\xi^2}.
\end{equation}

\begin{figure}[htp]
    \vspace{10pt}
    \centering
    \begin{tikzpicture}[
        scale=0.5,
        l/.style={draw=Black!30}, 
        s/.style={draw=Black, inner sep=0pt, minimum size=14pt}, 
        u/.style={ultra thick, draw=MyBlue, fill=MyBlue!10}, 
        b/.style={ultra thick, draw=MyPurple, fill=MyPurple!10}, 
        r/.style={ultra thick, draw=MyBlue}, 
        v/.style={ultra thick, draw=MyPurple}, 
        ]
        \draw[l] (-0.5, -0.5) grid (2.5, 4.5);
        \draw[l] (3.5, -0.5) grid (6.5, 4.5);
        \draw[l] (7.5, -0.5) grid (10.5, 4.5);
        \draw[l] (11.5, -0.5) grid (14.5, 4.5);
        \node[s] at (0.5, 0.5) {\scriptsize $n_{1}$};
        \node[s] at (0.5, 1.5) {\scriptsize $n_{2}$};
        \node[s] at (0.5, 2.5) {\scriptsize $n_{3}$};
        \node[s] at (0.5, 3.5) {\scriptsize $n_{4}$};
        \node[s, b] at (0.5, 4.5) {\scriptsize $n_{5}$};
        \node[s] at (5.5, 0.5) {\scriptsize $n_{1}$};
        \node[s] at (5.5, 2.5) {\scriptsize $n_{3}$};
        \node[s, u] at (5.5, 1.5) {\scriptsize $f_{2}$};
        \node[s, u] at (5.5, 3.5) {\scriptsize $f_{4}$};
        \node[s] at (8.5, 0.5) {\scriptsize $n_{1}$};
        \node[s] at (8.5, 1.5) {\scriptsize $n_{2}$};
        \node[s] at (8.5, 2.5) {\scriptsize $n_{3}$};
        \node[s] at (8.5, 3.5) {\scriptsize $n_{4}$};
        \node[s, b] at (8.5, -0.5) {\scriptsize $n_{0}$};
        \node[s] at (13.5, 1.5) {\scriptsize $n_{2}$};
        \node[s] at (13.5, 3.5) {\scriptsize $n_{4}$};
        \node[s, u] at (13.5, 0.5) {\scriptsize $f_{1}$};
        \node[s, u] at (13.5, 2.5) {\scriptsize $f_{3}$};
        \draw[r] (1.0, 0.5) -| (1.5, 1.5);
        \draw[r, ->, >=stealth, MyBlue] (1.0, 1.5) -- (5.0, 1.5);
        \draw[r] (1.0, 2.4) -| (1.5, 1.5);
        \draw[r] (1.0, 2.6) -| (1.5, 3.5);
        \draw[r, ->, >=stealth, MyBlue] (1.0, 3.5) -- (5.0, 3.5);
        \draw[r] (1.0, 4.5) -| (1.5, 3.5);
        \draw[r] (9.0, -0.5) -| (9.5, 0.5);
        \draw[r, ->, >=stealth, MyBlue] (9.0, 0.5) -- (13.0, 0.5);
        \draw[r] (9.0, 1.4) -| (9.5, 0.5);
        \draw[r] (9.0, 1.6) -| (9.5, 2.5);
        \draw[r, ->, >=stealth, MyBlue] (9.0, 2.5) -- (13.0, 2.5);
        \draw[r] (9.0, 3.5) -| (9.5, 2.5);
        \draw[v] (0.0, 0.5) -| (-0.5, 4.5);
        \draw[v] (0.0, 1.5) -- (-0.5, 1.5);
        \draw[v] (0.0, 2.5) -- (-0.5, 2.5);
        \draw[v] (0.0, 3.5) -- (-0.5, 3.5);
        \draw[v, ->, >=stealth, MyPurple] (-0.5, 4.5) -- (0.0, 4.5);
        \draw[v] (8.0, 3.5) -| (7.5, -0.5);
        \draw[v] (8.0, 2.5) -- (7.5, 2.5);
        \draw[v] (8.0, 1.5) -- (7.5, 1.5);
        \draw[v] (8.0, 0.5) -- (7.5, 0.5);
        \draw[v, ->, >=stealth, MyPurple] (7.5, -0.5) -- (8.0, -0.5);
        \draw[->, >=stealth] (2.2, 5.0) -- node[above] {\large $\RR$} (3.8, 5.0);
        \draw[->, >=stealth] (10.2, 5.0) -- node[above] {\large $\LL$} (11.8, 5.0);
        \draw[->, >=stealth] (-1.0, 1.0) -- node[left] {\large $i$} (-1.0, 3.0);
        \draw[->, >=stealth] (0.0, -1.0) -- node[below] {\large $t$} (2.0, -1.0);
    \end{tikzpicture}
    \caption{\textbf{Right and left boundary propagators.} The action of $\RR$ is equivalent to introducing an additional \emph{virtual} site on the top (represented by the purple square), initialize it in the state that depends on the four sites preceding it, and then evolving the second and fourth site according to the deterministic rule 201 (blue arrows). Similarly, the left boundary propagator $\LL$ can be reproduced by introducing a virtual site at the bottom, and then applying deterministic evolution.}
    \label{fig:stochasticBoundaries}
\end{figure}
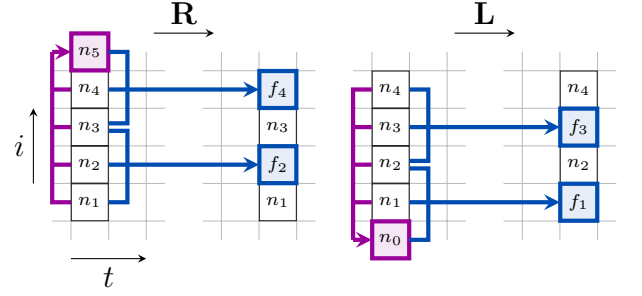

The boundary propagators~$\RR$ and $\LL$ are assumed to stochastically act only on the rightmost and leftmost sites respectively, while the other three sites change deterministically, according to the dynamical rule~\eqref{eq:rule-201}. Equivalently, we can imagine we temporarily introduce an additional site to the edge of the chain, in a state that depends on the configuration of the four sites, and update the site at the edge deterministically, as illustrated in Fig.~\ref{fig:stochasticBoundaries}. Explicitly, the matrix elements of~$\RR$ and~$\LL$ can be parametrized as
  \begin{equation}\label{eq:boundaryProps}
    \begin{aligned}
      R^{n_1^{\prime} n_2^{\prime} n_3^{\prime} n_4^{\prime}}_{n_1 n_2 n_3 n_4}
      &\!\!=\!
      \delta_{n_1^{\prime},n_1}
      \delta_{n_2^{\prime},f_2}
      \delta_{n_3^{\prime},n_3}
      \!\smashoperator{\sum_{n_5=0}^1}\!\delta_{n_4^{\prime},f_4} \phi^R_{n_1 n_2 n_3 n_4 n_5},\\
      L^{n_1^{\prime}n_2^{\prime} n_3^{\prime} n_4^{\prime}}_{n_1 n_2 n_3 n_4}
      &\!\!=\!
      \delta_{n_2^{\prime},n_2}
      \delta_{n_3^{\prime},f_3}
      \delta_{n_4^{\prime},n_4}
      \!\smashoperator{\sum_{n_0=0}^1}\!\delta_{n_1^{\prime},f_1} \phi^L_{n_0 n_1 n_2 n_3 n_4},
    \end{aligned}
  \end{equation}
where~$\phi^R_{n_1 n_2 n_3 n_4 n_5}$ and $\phi^{L}_{n_0 n_1 n_2 n_3 n_4}$ can be interpreted as conditional probabilities of the virtual sites being $n_5$ and $n_0$, respectively, if the configurations at the edge are~$(n_1 n_2 n_3 n_4)$. Here we use the shorthand notation~$f_i=f(n_{i-1},n_i,n_{i+1})$, as introduced in~\eqref{eq:local-update}. Additionally, the matrix elements in each column of $\RR$ and $\LL$ have to sum into $1$, which for any four-site configuration $(n_1 n_2 n_3 n_4)$ implies 
\begin{equation}\label{eq:normalization}
  \sum_{n_5=0}^1 \phi^R_{n_1 n_2 n_3 n_4 n_5}=
  \sum_{n_0=0}^1 \phi^L_{n_0 n_1 n_2 n_3 n_4}=1.
\end{equation}

Applying the dynamical rule~\eqref{eq:rule-201} to the ansatz~\eqref{eq:boundaryProps} while taking into account the normalization condition~\eqref{eq:normalization} it immediately follows that for any combination of $n_1,n_2,n_3,n_4$ the following holds
\begin{equation}
  R^{n_1 n_2 1 n_4}_{n_1 n_2 1 n_4}=
  L^{n_1 1 n_3 n_4}_{n_1 1 n_3 n_4}=1.
\end{equation}
Furthermore, we note that the steady state is restricted to the subspace without pairs of~$1$, therefore we can without loss of generality set
\begin{equation}
  R^{1100}_{1100}=R^{1101}_{1101}=L^{0011}_{0011}=L^{1011}_{1011}=1.
\end{equation}
After reducing the number of parameters, we are left with $3$ non-deterministic $2\times 2$ blocks per boundary propagator, each one of them given by two parameters, either $(\phi^R_{n_1 n_2 n_3 01},\phi^R_{n_1 n_2 n_3 11})$ or~$(\phi^L_{10 n_1 n_2 n_3},\phi^L_{11 n_1 n_2 n_3})$, with the fixed configuration~$(n_1, n_2, n_3)$. Plugging the ansatz into boundary equations~\eqref{eq:boundaryAlgebra1} reduces the number of parameters to one per block. Explicitly,
\begin{equation}\label{eq:generalSol1}
  \begin{aligned}
  \phi^{R}_{00001}&=
    \frac{\omega\big((\lambda-1)\omega+\xi^2\big)}
    {\lambda\big((\lambda-1)\xi+\omega^2\big)}+
  \theta^{R}_{1},\\
  \phi^{R}_{00011}&=
    \frac{(\lambda-1)\xi+\omega^2}
    {\xi\big( (\lambda-1)^2-\xi\omega\big)}\theta^R_1,\\
  \phi^{R}_{01001}&=
    \frac{\xi\Big((\lambda-1)\omega+\xi^2\big)}
    {\lambda(\lambda-1)\big((\lambda-1)^2-\xi\omega\big)}+
  \theta^{R}_2,\\
  \phi^{R}_{01011}&=(\lambda-1)\theta^{R}_2,\\
  \phi^{R}_{10001}&=
    \frac{\omega\big((\lambda-1)\omega+\xi^2\big)}
    {\lambda\big((\lambda-1)\xi+\omega^2\big)}+
  \theta^{R}_{3},\\
  \phi^{R}_{10011}&=
    \frac{\xi\Big((\lambda-1)\xi+\omega^2\Big)}
    {\omega\Big((\lambda-1)^2-\xi \omega\Big)}\theta^{R}_3,
  \end{aligned}
\end{equation}
where~$\theta_{1,2,3}^R$ are the free parameters corresponding to the three non-deterministic blocks. Analogously, introducing the left-boundary coefficients~$\theta_{1,2,3}^{L}$, the solution to~\eqref{eq:boundaryAlgebra2} is given by,
\begin{equation}\label{eq:generalSol2}
  \begin{aligned}
  \phi^{L}_{10000}&=
    \frac{\xi\big((\lambda-1)\xi+\omega^2\big)}
    {\lambda\big((\lambda-1)\omega+\xi^2\big)}+
  \theta^{L}_{1},\\
  \phi^{L}_{11000}&=
    \frac{(\lambda-1)\omega+\xi^2}
    {\omega\big( (\lambda-1)^2-\xi\omega\big)}\theta^{L}_1,\\
  \phi^{L}_{10010}&=
    \frac{\omega\Big( (\lambda-1)\xi+\omega^2\big)}
    {\lambda(\lambda-1)\big((\lambda-1)^2-\xi\omega\big)}+
  \theta^{L}_2,\\
  \phi^{L}_{11010}&=(\lambda-1)\theta^{L}_2,\\
  \phi^{L}_{10001}&=\frac{\xi\big((\lambda-1)\xi+\omega^2\big)}
    {\lambda\big((\lambda-1)\omega+\xi^2\big)}+
  \theta^{L}_{3},\\
  \phi^{L}_{11001}&=\frac{\omega\Big( (\lambda-1)\omega+\xi^2\Big)}
    {\xi\Big( (\lambda-1)^2-\xi\omega\Big)}\theta^{L}_3.
  \end{aligned}
\end{equation}
Equations~\eqref{eq:generalSol1} and~\eqref{eq:generalSol2} provide the most general form of the boundary propagators~$\RR$ and~$\LL$, for which the asymptotic state introduced in the previous subsection is the fixed point. Note that the parameters~$\theta_{1,2,3}^{R/L}$ are not completely arbitrary, since all the matrix elements of the stochastic boundary matrices should be between $0$ and $1$.

A particularly convenient choice of parametrization is to set~$\theta_{1,2,3}^{R/L}=0$. In this case the stochastic blocks can be summarized by
\begin{equation}
  \begin{aligned}
  \phi_{n_1 n_2 n_3 n_4 n_5}^R &=
  \frac{p_{n_1 n_2 n_3 n_4 n_5 0}+p_{n_1 n_2 n_3 n_4 n_5 1}}
  {p_{n_1 n_2 n_3 n_4}},\\
  \phi_{n_0 n_1 n_2 n_3 n_4}^L &=
  \frac{p^{\prime}_{0 n_0 n_1 n_2 n_3 n_4}+p^{\prime}_{1 n_0 n_1 n_2 n_3 n_4}}
  {p^{\prime}_{n_1 n_2 n_3 n_4}}.
  \end{aligned}
\end{equation}
This is reminiscent of the situation observed in RCA54 (see e.g.\ \cite{Klobas2019b}): if the $4$ spins at the edge are in the configuration $(n_1 n_2 n_3 n_4)$, the probability of finding the virtual site to the right (or left) in the state $n_5$ (or $n_0$) is the same as the conditional Gibbs probability of observing the~$5$-site configuration, given the knowledge of the state of the first $4$ sites. The construction proves that the equilibrium distribution of finite configurations can be equivalently understood as a steady state of a boundary-driven system. Note that this does not apply to \emph{dynamics}. Starting with a configuration on a finite subsection of the periodic lattice, while assuming a random distribution elsewhere (as described in Subsec.~\ref{subsec:asymptDistr}), evolving it in time and at the end averaging over all the sites outside of the finite subsection we started with, will give us a different distribution compared to taking the same initial configuration and evolving it with the stochastic boundaries.

The construction in this section represents a class of non-trivial boundary propagators, for which the NESS is particularly simple. Generalizing boundary vectors to encode the information about the sites close to the boundary (similar to the situation considered in~\cite{Prosen2017,Buca2019}), might provide a richer family of stochastic boundary propagators with nontrivial NESS. However, this is beyond the scope of this paper and the full classification of all possible solvable (or integrable) boundaries remains an open question.

\section{\label{sec:conclusion} Conclusions}
In this paper we have studied in detail the dynamics of the RCA201/Floquet-PXP model, a classical deterministic reversible cellular automaton. This model is to the classical PXP model (or one-dimensional two-spin facilitated FA model) what the RCA54 is to the classical stochastic FA model: a deterministic lattice system with periodic circuit-dynamics with the same kinetic constraint of the corresponding KCM. The study of these RCAs thus allows us to extend our understanding of the consequences of constraints to dynamics. 

We have shown that the RCA201/Floquet-PXP model is integrable. Its dynamics is fully determined by conserved quasiparticles that propagate ballistically and interact via collisions. As usual, integrability implies that many properties of the model can be investigated exactly. Most notably, we have calculated the exact non-equilibrium stationary state, which takes the form of a low bond dimension MPS, under both periodic and stochastic boundary conditions. The methods we applied are similar to those employed to solve the RCA54 model. Note however that the RCA201 is a slightly more complicated model. In particular, the stricter kinetic constraint forces the dynamics to be always strictly out of equilibrium due to the underlying period three cycling of its three-fold vacua (which implies the existence of probability currents under all conditions).

Our work here opens the door for obtaining several other exact results for the dynamics of the RCA201/Floquet-PXP model, just like it was done recently for the RCA54. We anticipate the following: (i) the exact large deviation statistics of trajectory observables, cf.~\cite{Buca2019}; (ii) the exact MPS form of the ``time state'', that is, the probability vector that encodes all time-correlators that are local in space, cf.~\cite{Klobas2019b}; (iii) construction of the MPS representation for the time-evolution of local observables and the explicit solutions of the dynamical correlation functions and quench dynamics, cf.~\cite{Klobas2019}; (iv) the properties of the dual system to the RCA201 where propagation is in the space rather than time direction, and the consequences of this duality, cf.~\cite{Klobas2020}. We hope to report on some of these in the near future.

\bigskip

\acknowledgments

We acknowledge support of The Leverhulme Trust through Grant number RPG-2018-181. KK and TP acknowledge support from European Research Council (ERC) under Advanced Grant 694544 -- OMNES and the program P1-0402 of Slovenian Research Agency.

\bibliography{sections/bibfile}

\begin{thebibliography}{46}%
\makeatletter
\providecommand \@ifxundefined [1]{%
 \@ifx{#1\undefined}
}%
\providecommand \@ifnum [1]{%
 \ifnum #1\expandafter \@firstoftwo
 \else \expandafter \@secondoftwo
 \fi
}%
\providecommand \@ifx [1]{%
 \ifx #1\expandafter \@firstoftwo
 \else \expandafter \@secondoftwo
 \fi
}%
\providecommand \natexlab [1]{#1}%
\providecommand \enquote  [1]{``#1''}%
\providecommand \bibnamefont  [1]{#1}%
\providecommand \bibfnamefont [1]{#1}%
\providecommand \citenamefont [1]{#1}%
\providecommand \href@noop [0]{\@secondoftwo}%
\providecommand \href [0]{\begingroup \@sanitize@url \@href}%
\providecommand \@href[1]{\@@startlink{#1}\@@href}%
\providecommand \@@href[1]{\endgroup#1\@@endlink}%
\providecommand \@sanitize@url [0]{\catcode `\\12\catcode `\$12\catcode
  `\&12\catcode `\#12\catcode `\^12\catcode `\_12\catcode `\%12\relax}%
\providecommand \@@startlink[1]{}%
\providecommand \@@endlink[0]{}%
\providecommand \url  [0]{\begingroup\@sanitize@url \@url }%
\providecommand \@url [1]{\endgroup\@href {#1}{\urlprefix }}%
\providecommand \urlprefix  [0]{URL }%
\providecommand \Eprint [0]{\href }%
\providecommand \doibase [0]{https://doi.org/}%
\providecommand \selectlanguage [0]{\@gobble}%
\providecommand \bibinfo  [0]{\@secondoftwo}%
\providecommand \bibfield  [0]{\@secondoftwo}%
\providecommand \translation [1]{[#1]}%
\providecommand \BibitemOpen [0]{}%
\providecommand \bibitemStop [0]{}%
\providecommand \bibitemNoStop [0]{.\EOS\space}%
\providecommand \EOS [0]{\spacefactor3000\relax}%
\providecommand \BibitemShut  [1]{\csname bibitem#1\endcsname}%
\let\auto@bib@innerbib\@empty
\bibitem [{\citenamefont {Bobenko}\ \emph {et~al.}(1993)\citenamefont
  {Bobenko}, \citenamefont {Bordemann}, \citenamefont {Gunn},\ and\
  \citenamefont {Pinkall}}]{Bobenko1993}%
  \BibitemOpen
  \bibfield  {author} {\bibinfo {author} {\bibfnamefont {A.}~\bibnamefont
  {Bobenko}}, \bibinfo {author} {\bibfnamefont {M.}~\bibnamefont {Bordemann}},
  \bibinfo {author} {\bibfnamefont {C.}~\bibnamefont {Gunn}},\ and\ \bibinfo
  {author} {\bibfnamefont {U.}~\bibnamefont {Pinkall}},\ }\bibfield  {title}
  {\bibinfo {title} {{On two integrable cellular automata}},\ }\href
  {https://doi.org/10.1007/BF02097234} {\bibfield  {journal} {\bibinfo
  {journal} {Commun. Math. Phys.}\ }\textbf {\bibinfo {volume} {158}},\
  \bibinfo {pages} {127} (\bibinfo {year} {1993})}\BibitemShut {NoStop}%
\bibitem [{\citenamefont {Fisher}(1961)}]{Fisher1961}%
  \BibitemOpen
  \bibfield  {author} {\bibinfo {author} {\bibfnamefont {M.~E.}\ \bibnamefont
  {Fisher}},\ }\bibfield  {title} {\bibinfo {title} {Statistical mechanics of
  dimers on a plane lattice},\ }\href
  {https://doi.org/10.1103/PhysRev.124.1664} {\bibfield  {journal} {\bibinfo
  {journal} {Phys. Rev.}\ }\textbf {\bibinfo {volume} {124}},\ \bibinfo {pages}
  {1664} (\bibinfo {year} {1961})}\BibitemShut {NoStop}%
\bibitem [{\citenamefont {Henley}(2010)}]{Henley2010}%
  \BibitemOpen
  \bibfield  {author} {\bibinfo {author} {\bibfnamefont {C.~L.}\ \bibnamefont
  {Henley}},\ }\bibfield  {title} {\bibinfo {title} {The ``coulomb phase'' in
  frustrated systems},\ }\href
  {https://doi.org/10.1146/annurev-conmatphys-070909-104138} {\bibfield
  {journal} {\bibinfo  {journal} {Annu. Rev. Condens. Matter Phys.}\ }\textbf
  {\bibinfo {volume} {1}},\ \bibinfo {pages} {179} (\bibinfo {year}
  {2010})}\BibitemShut {NoStop}%
\bibitem [{\citenamefont {Moessner}\ and\ \citenamefont
  {Raman}(2011)}]{Moessner2011}%
  \BibitemOpen
  \bibfield  {author} {\bibinfo {author} {\bibfnamefont {R.}~\bibnamefont
  {Moessner}}\ and\ \bibinfo {author} {\bibfnamefont {K.}~\bibnamefont
  {Raman}},\ }\bibfield  {title} {\bibinfo {title} {Quantum dimer models},\
  }in\ \href@noop {} {\emph {\bibinfo {booktitle} {Introduction to frustrated
  magnetism: materials, experiments, theory}}},\ Vol.\ \bibinfo {volume}
  {164},\ \bibinfo {editor} {edited by\ \bibinfo {editor} {\bibfnamefont
  {C.}~\bibnamefont {Lacroix}}, \bibinfo {editor} {\bibfnamefont
  {P.}~\bibnamefont {Mendels}},\ and\ \bibinfo {editor} {\bibfnamefont
  {F.}~\bibnamefont {Mila}}}\ (\bibinfo  {publisher} {Springer Science \&
  Business Media},\ \bibinfo {year} {2011})\ Chap.~\bibinfo {chapter} {17},
  pp.\ \bibinfo {pages} {437--477}\BibitemShut {NoStop}%
\bibitem [{\citenamefont {Chalker}(2017)}]{Chalker2014}%
  \BibitemOpen
  \bibfield  {author} {\bibinfo {author} {\bibfnamefont {J.~T.}\ \bibnamefont
  {Chalker}},\ }\bibfield  {title} {\bibinfo {title} {Spin liquids and
  frustrated magnetism},\ }in\ \href@noop {} {\emph {\bibinfo {booktitle}
  {Topological Aspects of Condensed Matter Physics: Lecture Notes of the Les
  Houches Summer School: Volume 103, August 2014}}},\ Vol.\ \bibinfo {volume}
  {103},\ \bibinfo {editor} {edited by\ \bibinfo {editor} {\bibfnamefont
  {C.}~\bibnamefont {Chamon}}, \bibinfo {editor} {\bibfnamefont {M.~O.}\
  \bibnamefont {Goerbig}}, \bibinfo {editor} {\bibfnamefont {R.}~\bibnamefont
  {Moessner}},\ and\ \bibinfo {editor} {\bibfnamefont {L.~F.}\ \bibnamefont
  {Cugliandolo}}}\ (\bibinfo  {publisher} {Oxford University Press},\ \bibinfo
  {year} {2017})\ Chap.~\bibinfo {chapter} {3}, pp.\ \bibinfo {pages}
  {123--162}\BibitemShut {NoStop}%
\bibitem [{\citenamefont {Fredrickson}\ and\ \citenamefont
  {Andersen}(1984)}]{Fredrickson1984}%
  \BibitemOpen
  \bibfield  {author} {\bibinfo {author} {\bibfnamefont {G.~H.}\ \bibnamefont
  {Fredrickson}}\ and\ \bibinfo {author} {\bibfnamefont {H.~C.}\ \bibnamefont
  {Andersen}},\ }\bibfield  {title} {\bibinfo {title} {Kinetic ising model of
  the glass transition},\ }\href {https://doi.org/10.1103/PhysRevLett.53.1244}
  {\bibfield  {journal} {\bibinfo  {journal} {Phys. Rev. Lett.}\ }\textbf
  {\bibinfo {volume} {53}},\ \bibinfo {pages} {1244} (\bibinfo {year}
  {1984})}\BibitemShut {NoStop}%
\bibitem [{\citenamefont {Palmer}\ \emph {et~al.}(1984)\citenamefont {Palmer},
  \citenamefont {Stein}, \citenamefont {Abrahams},\ and\ \citenamefont
  {Anderson}}]{Palmer1984}%
  \BibitemOpen
  \bibfield  {author} {\bibinfo {author} {\bibfnamefont {R.~G.}\ \bibnamefont
  {Palmer}}, \bibinfo {author} {\bibfnamefont {D.~L.}\ \bibnamefont {Stein}},
  \bibinfo {author} {\bibfnamefont {E.}~\bibnamefont {Abrahams}},\ and\
  \bibinfo {author} {\bibfnamefont {P.~W.}\ \bibnamefont {Anderson}},\
  }\bibfield  {title} {\bibinfo {title} {Models of hierarchically constrained
  dynamics for glassy relaxation},\ }\href
  {https://doi.org/10.1103/PhysRevLett.53.958} {\bibfield  {journal} {\bibinfo
  {journal} {Phys. Rev. Lett.}\ }\textbf {\bibinfo {volume} {53}},\ \bibinfo
  {pages} {958} (\bibinfo {year} {1984})}\BibitemShut {NoStop}%
\bibitem [{\citenamefont {J{\"a}ckle}\ and\ \citenamefont
  {Eisinger}(1991)}]{Jackle1991}%
  \BibitemOpen
  \bibfield  {author} {\bibinfo {author} {\bibfnamefont {J.}~\bibnamefont
  {J{\"a}ckle}}\ and\ \bibinfo {author} {\bibfnamefont {S.~Z.}\ \bibnamefont
  {Eisinger}},\ }\bibfield  {title} {\bibinfo {title} {A hierarchically
  constrained kinetic ising model},\ }\href
  {https://doi.org/10.1007/BF01453764} {\bibfield  {journal} {\bibinfo
  {journal} {Z. fur Phys. B}\ }\textbf {\bibinfo {volume} {84}},\ \bibinfo
  {pages} {115} (\bibinfo {year} {1991})}\BibitemShut {NoStop}%
\bibitem [{\citenamefont {Ritort}\ and\ \citenamefont
  {Sollich}(2003)}]{Ritort2003}%
  \BibitemOpen
  \bibfield  {author} {\bibinfo {author} {\bibfnamefont {F.}~\bibnamefont
  {Ritort}}\ and\ \bibinfo {author} {\bibfnamefont {P.}~\bibnamefont
  {Sollich}},\ }\bibfield  {title} {\bibinfo {title} {Glassy dynamics of
  kinetically constrained models},\ }\href
  {https://doi.org/10.1080/0001873031000093582} {\bibfield  {journal} {\bibinfo
   {journal} {Adv. Phys.}\ }\textbf {\bibinfo {volume} {52}},\ \bibinfo {pages}
  {219} (\bibinfo {year} {2003})}\BibitemShut {NoStop}%
\bibitem [{\citenamefont {Garrahan}\ \emph {et~al.}(2011)\citenamefont
  {Garrahan}, \citenamefont {Sollich},\ and\ \citenamefont
  {Toninelli}}]{Garrahan2011}%
  \BibitemOpen
  \bibfield  {author} {\bibinfo {author} {\bibfnamefont {J.~P.}\ \bibnamefont
  {Garrahan}}, \bibinfo {author} {\bibfnamefont {P.}~\bibnamefont {Sollich}},\
  and\ \bibinfo {author} {\bibfnamefont {C.}~\bibnamefont {Toninelli}},\
  }\bibfield  {title} {\bibinfo {title} {{Kinetically Constrained Models}},\
  }in\ \href@noop {} {\emph {\bibinfo {booktitle} {Dynamical Heterogeneities in
  Glasses, Colloids, and Granular Media}}},\ \bibinfo {series and number}
  {International Series of Monographs on Physics},\ \bibinfo {editor} {edited
  by\ \bibinfo {editor} {\bibfnamefont {L.}~\bibnamefont {Berthier}}, \bibinfo
  {editor} {\bibfnamefont {G.}~\bibnamefont {Biroli}}, \bibinfo {editor}
  {\bibfnamefont {J.-P.}\ \bibnamefont {Bouchaud}}, \bibinfo {editor}
  {\bibfnamefont {L.}~\bibnamefont {Cipelletti}},\ and\ \bibinfo {editor}
  {\bibfnamefont {W.}~\bibnamefont {van Saarloos}}}\ (\bibinfo  {publisher}
  {Oxford University Press},\ \bibinfo {address} {Oxford, UK},\ \bibinfo {year}
  {2011})\ Chap.~\bibinfo {chapter} {10}, pp.\ \bibinfo {pages}
  {341--366}\BibitemShut {NoStop}%
\bibitem [{\citenamefont {Garrahan}(2018)}]{Garrahan2018}%
  \BibitemOpen
  \bibfield  {author} {\bibinfo {author} {\bibfnamefont {J.~P.}\ \bibnamefont
  {Garrahan}},\ }\bibfield  {title} {\bibinfo {title} {Aspects of
  non-equilibrium in classical and quantum systems: Slow relaxation and
  glasses, dynamical large deviations, quantum non-ergodicity, and open quantum
  dynamics},\ }\href
  {https://doi.org/https://doi.org/10.1016/j.physa.2017.12.149} {\bibfield
  {journal} {\bibinfo  {journal} {Physica A}\ }\textbf {\bibinfo {volume}
  {504}},\ \bibinfo {pages} {130} (\bibinfo {year} {2018})}\BibitemShut
  {NoStop}%
\bibitem [{\citenamefont {van Horssen}\ \emph {et~al.}(2015)\citenamefont {van
  Horssen}, \citenamefont {Levi},\ and\ \citenamefont
  {Garrahan}}]{Horssen2015}%
  \BibitemOpen
  \bibfield  {author} {\bibinfo {author} {\bibfnamefont {M.}~\bibnamefont {van
  Horssen}}, \bibinfo {author} {\bibfnamefont {E.}~\bibnamefont {Levi}},\ and\
  \bibinfo {author} {\bibfnamefont {J.~P.}\ \bibnamefont {Garrahan}},\
  }\bibfield  {title} {\bibinfo {title} {Dynamics of many-body localization in
  a translation-invariant quantum glass model},\ }\href
  {https://doi.org/10.1103/PhysRevB.92.100305} {\bibfield  {journal} {\bibinfo
  {journal} {Phys. Rev. B}\ }\textbf {\bibinfo {volume} {92}},\ \bibinfo
  {pages} {100305} (\bibinfo {year} {2015})}\BibitemShut {NoStop}%
\bibitem [{\citenamefont {Lan}\ \emph {et~al.}(2018)\citenamefont {Lan},
  \citenamefont {van Horssen}, \citenamefont {Powell},\ and\ \citenamefont
  {Garrahan}}]{Lan2018}%
  \BibitemOpen
  \bibfield  {author} {\bibinfo {author} {\bibfnamefont {Z.}~\bibnamefont
  {Lan}}, \bibinfo {author} {\bibfnamefont {M.}~\bibnamefont {van Horssen}},
  \bibinfo {author} {\bibfnamefont {S.}~\bibnamefont {Powell}},\ and\ \bibinfo
  {author} {\bibfnamefont {J.~P.}\ \bibnamefont {Garrahan}},\ }\bibfield
  {title} {\bibinfo {title} {Quantum slow relaxation and metastability due to
  dynamical constraints},\ }\href
  {https://doi.org/10.1103/PhysRevLett.121.040603} {\bibfield  {journal}
  {\bibinfo  {journal} {Phys. Rev. Lett.}\ }\textbf {\bibinfo {volume} {121}},\
  \bibinfo {pages} {040603} (\bibinfo {year} {2018})}\BibitemShut {NoStop}%
\bibitem [{\citenamefont {Lesanovsky}(2011)}]{Lesanovsky2011}%
  \BibitemOpen
  \bibfield  {author} {\bibinfo {author} {\bibfnamefont {I.}~\bibnamefont
  {Lesanovsky}},\ }\bibfield  {title} {\bibinfo {title} {Many-body spin
  interactions and the ground state of a dense rydberg lattice gas},\ }\href
  {https://doi.org/10.1103/PhysRevLett.106.025301} {\bibfield  {journal}
  {\bibinfo  {journal} {Phys. Rev. Lett.}\ }\textbf {\bibinfo {volume} {106}},\
  \bibinfo {pages} {025301} (\bibinfo {year} {2011})}\BibitemShut {NoStop}%
\bibitem [{\citenamefont {Turner}\ \emph {et~al.}(2018)\citenamefont {Turner},
  \citenamefont {Michailidis}, \citenamefont {Abanin}, \citenamefont {Serbyn},\
  and\ \citenamefont {Papi{\'c}}}]{Turner2018}%
  \BibitemOpen
  \bibfield  {author} {\bibinfo {author} {\bibfnamefont {C.~J.}\ \bibnamefont
  {Turner}}, \bibinfo {author} {\bibfnamefont {A.~A.}\ \bibnamefont
  {Michailidis}}, \bibinfo {author} {\bibfnamefont {D.~A.}\ \bibnamefont
  {Abanin}}, \bibinfo {author} {\bibfnamefont {M.}~\bibnamefont {Serbyn}},\
  and\ \bibinfo {author} {\bibfnamefont {Z.}~\bibnamefont {Papi{\'c}}},\
  }\bibfield  {title} {\bibinfo {title} {Weak ergodicity breaking from quantum
  many-body scars},\ }\href {https://doi.org/10.1038/s41567-018-0137-5}
  {\bibfield  {journal} {\bibinfo  {journal} {Nature Physics}\ }\textbf
  {\bibinfo {volume} {14}},\ \bibinfo {pages} {745} (\bibinfo {year}
  {2018})}\BibitemShut {NoStop}%
\bibitem [{\citenamefont {Pancotti}\ \emph {et~al.}(2020)\citenamefont
  {Pancotti}, \citenamefont {Giudice}, \citenamefont {Cirac}, \citenamefont
  {Garrahan},\ and\ \citenamefont {Ba\~nuls}}]{Pancotti2020}%
  \BibitemOpen
  \bibfield  {author} {\bibinfo {author} {\bibfnamefont {N.}~\bibnamefont
  {Pancotti}}, \bibinfo {author} {\bibfnamefont {G.}~\bibnamefont {Giudice}},
  \bibinfo {author} {\bibfnamefont {J.~I.}\ \bibnamefont {Cirac}}, \bibinfo
  {author} {\bibfnamefont {J.~P.}\ \bibnamefont {Garrahan}},\ and\ \bibinfo
  {author} {\bibfnamefont {M.~C.}\ \bibnamefont {Ba\~nuls}},\ }\bibfield
  {title} {\bibinfo {title} {Quantum east model: Localization, nonthermal
  eigenstates, and slow dynamics},\ }\href
  {https://doi.org/10.1103/PhysRevX.10.021051} {\bibfield  {journal} {\bibinfo
  {journal} {Phys. Rev. X}\ }\textbf {\bibinfo {volume} {10}},\ \bibinfo
  {pages} {021051} (\bibinfo {year} {2020})}\BibitemShut {NoStop}%
\bibitem [{\citenamefont {Nahum}\ \emph {et~al.}(2017)\citenamefont {Nahum},
  \citenamefont {Ruhman}, \citenamefont {Vijay},\ and\ \citenamefont
  {Haah}}]{Nahum1}%
  \BibitemOpen
  \bibfield  {author} {\bibinfo {author} {\bibfnamefont {A.}~\bibnamefont
  {Nahum}}, \bibinfo {author} {\bibfnamefont {J.}~\bibnamefont {Ruhman}},
  \bibinfo {author} {\bibfnamefont {S.}~\bibnamefont {Vijay}},\ and\ \bibinfo
  {author} {\bibfnamefont {J.}~\bibnamefont {Haah}},\ }\bibfield  {title}
  {\bibinfo {title} {Quantum entanglement growth under random unitary
  dynamics},\ }\href {https://doi.org/10.1103/PhysRevX.7.031016} {\bibfield
  {journal} {\bibinfo  {journal} {Phys. Rev. X}\ }\textbf {\bibinfo {volume}
  {7}},\ \bibinfo {pages} {031016} (\bibinfo {year} {2017})}\BibitemShut
  {NoStop}%
\bibitem [{\citenamefont {Nahum}\ \emph {et~al.}(2018)\citenamefont {Nahum},
  \citenamefont {Vijay},\ and\ \citenamefont {Haah}}]{Nahum2}%
  \BibitemOpen
  \bibfield  {author} {\bibinfo {author} {\bibfnamefont {A.}~\bibnamefont
  {Nahum}}, \bibinfo {author} {\bibfnamefont {S.}~\bibnamefont {Vijay}},\ and\
  \bibinfo {author} {\bibfnamefont {J.}~\bibnamefont {Haah}},\ }\bibfield
  {title} {\bibinfo {title} {Operator spreading in random unitary circuits},\
  }\href {https://doi.org/10.1103/PhysRevX.8.021014} {\bibfield  {journal}
  {\bibinfo  {journal} {Phys. Rev. X}\ }\textbf {\bibinfo {volume} {8}},\
  \bibinfo {pages} {021014} (\bibinfo {year} {2018})}\BibitemShut {NoStop}%
\bibitem [{\citenamefont {Chan}\ \emph {et~al.}(2018)\citenamefont {Chan},
  \citenamefont {De~Luca},\ and\ \citenamefont {Chalker}}]{DeLuca}%
  \BibitemOpen
  \bibfield  {author} {\bibinfo {author} {\bibfnamefont {A.}~\bibnamefont
  {Chan}}, \bibinfo {author} {\bibfnamefont {A.}~\bibnamefont {De~Luca}},\ and\
  \bibinfo {author} {\bibfnamefont {J.}~\bibnamefont {Chalker}},\ }\bibfield
  {title} {\bibinfo {title} {Solution of a minimal model for many-body quantum
  chaos},\ }\href {https://doi.org/10.1103/PhysRevX.8.041019} {\bibfield
  {journal} {\bibinfo  {journal} {Phys. Rev. X}\ }\textbf {\bibinfo {volume}
  {8}},\ \bibinfo {pages} {041019} (\bibinfo {year} {2018})}\BibitemShut
  {NoStop}%
\bibitem [{\citenamefont {Bertini}\ \emph {et~al.}(2019)\citenamefont
  {Bertini}, \citenamefont {Kos},\ and\ \citenamefont {Prosen}}]{Bertini2019}%
  \BibitemOpen
  \bibfield  {author} {\bibinfo {author} {\bibfnamefont {B.}~\bibnamefont
  {Bertini}}, \bibinfo {author} {\bibfnamefont {P.}~\bibnamefont {Kos}},\ and\
  \bibinfo {author} {\bibfnamefont {T.}~\bibnamefont {Prosen}},\ }\bibfield
  {title} {\bibinfo {title} {Exact correlation functions for dual-unitary
  lattice models in 1+ 1 dimensions},\ }\href
  {https://doi.org/10.1103/PhysRevLett.123.210601} {\bibfield  {journal}
  {\bibinfo  {journal} {Phys. Rev. Lett.}\ }\textbf {\bibinfo {volume} {123}},\
  \bibinfo {pages} {210601} (\bibinfo {year} {2019})}\BibitemShut {NoStop}%
\bibitem [{\citenamefont {Von~Keyserlingk}\ \emph {et~al.}(2018)\citenamefont
  {Von~Keyserlingk}, \citenamefont {Rakovszky}, \citenamefont {Pollmann},\ and\
  \citenamefont {Sondhi}}]{Pollmann}%
  \BibitemOpen
  \bibfield  {author} {\bibinfo {author} {\bibfnamefont {C.}~\bibnamefont
  {Von~Keyserlingk}}, \bibinfo {author} {\bibfnamefont {T.}~\bibnamefont
  {Rakovszky}}, \bibinfo {author} {\bibfnamefont {F.}~\bibnamefont
  {Pollmann}},\ and\ \bibinfo {author} {\bibfnamefont {S.~L.}\ \bibnamefont
  {Sondhi}},\ }\bibfield  {title} {\bibinfo {title} {Operator hydrodynamics,
  otocs, and entanglement growth in systems without conservation laws},\ }\href
  {https://doi.org/10.1103/PhysRevX.8.021013} {\bibfield  {journal} {\bibinfo
  {journal} {Phys. Rev. X}\ }\textbf {\bibinfo {volume} {8}},\ \bibinfo {pages}
  {021013} (\bibinfo {year} {2018})}\BibitemShut {NoStop}%
\bibitem [{\citenamefont {Rakovszky}\ \emph {et~al.}(2018)\citenamefont
  {Rakovszky}, \citenamefont {Pollmann},\ and\ \citenamefont {von
  Keyserlingk}}]{Pollman2}%
  \BibitemOpen
  \bibfield  {author} {\bibinfo {author} {\bibfnamefont {T.}~\bibnamefont
  {Rakovszky}}, \bibinfo {author} {\bibfnamefont {F.}~\bibnamefont
  {Pollmann}},\ and\ \bibinfo {author} {\bibfnamefont {C.~W.}\ \bibnamefont
  {von Keyserlingk}},\ }\bibfield  {title} {\bibinfo {title} {Diffusive
  hydrodynamics of out-of-time-ordered correlators with charge conservation},\
  }\href {https://doi.org/10.1103/PhysRevX.8.031058} {\bibfield  {journal}
  {\bibinfo  {journal} {Phys. Rev. X}\ }\textbf {\bibinfo {volume} {8}},\
  \bibinfo {pages} {031058} (\bibinfo {year} {2018})}\BibitemShut {NoStop}%
\bibitem [{\citenamefont {S\"underhauf}\ \emph {et~al.}(2018)\citenamefont
  {S\"underhauf}, \citenamefont {P\'erez-Garc\'{\i}a}, \citenamefont {Huse},
  \citenamefont {Schuch},\ and\ \citenamefont {Cirac}}]{sunderhauf}%
  \BibitemOpen
  \bibfield  {author} {\bibinfo {author} {\bibfnamefont {C.}~\bibnamefont
  {S\"underhauf}}, \bibinfo {author} {\bibfnamefont {D.}~\bibnamefont
  {P\'erez-Garc\'{\i}a}}, \bibinfo {author} {\bibfnamefont {D.~A.}\
  \bibnamefont {Huse}}, \bibinfo {author} {\bibfnamefont {N.}~\bibnamefont
  {Schuch}},\ and\ \bibinfo {author} {\bibfnamefont {J.~I.}\ \bibnamefont
  {Cirac}},\ }\bibfield  {title} {\bibinfo {title} {Localization with random
  time-periodic quantum circuits},\ }\href
  {https://doi.org/10.1103/PhysRevB.98.134204} {\bibfield  {journal} {\bibinfo
  {journal} {Phys. Rev. B}\ }\textbf {\bibinfo {volume} {98}},\ \bibinfo
  {pages} {134204} (\bibinfo {year} {2018})}\BibitemShut {NoStop}%
\bibitem [{\citenamefont {Khemani}\ \emph {et~al.}(2018)\citenamefont
  {Khemani}, \citenamefont {Vishwanath},\ and\ \citenamefont {Huse}}]{Khemani}%
  \BibitemOpen
  \bibfield  {author} {\bibinfo {author} {\bibfnamefont {V.}~\bibnamefont
  {Khemani}}, \bibinfo {author} {\bibfnamefont {A.}~\bibnamefont
  {Vishwanath}},\ and\ \bibinfo {author} {\bibfnamefont {D.~A.}\ \bibnamefont
  {Huse}},\ }\bibfield  {title} {\bibinfo {title} {Operator spreading and the
  emergence of dissipative hydrodynamics under unitary evolution with
  conservation laws},\ }\href {https://doi.org/10.1103/PhysRevX.8.031057}
  {\bibfield  {journal} {\bibinfo  {journal} {Phys. Rev. X}\ }\textbf {\bibinfo
  {volume} {8}},\ \bibinfo {pages} {031057} (\bibinfo {year}
  {2018})}\BibitemShut {NoStop}%
\bibitem [{\citenamefont {Pai}\ \emph {et~al.}(2019)\citenamefont {Pai},
  \citenamefont {Pretko},\ and\ \citenamefont {Nandkishore}}]{Pretko}%
  \BibitemOpen
  \bibfield  {author} {\bibinfo {author} {\bibfnamefont {S.}~\bibnamefont
  {Pai}}, \bibinfo {author} {\bibfnamefont {M.}~\bibnamefont {Pretko}},\ and\
  \bibinfo {author} {\bibfnamefont {R.~M.}\ \bibnamefont {Nandkishore}},\
  }\bibfield  {title} {\bibinfo {title} {Localization in fractonic random
  circuits},\ }\href {https://doi.org/10.1103/PhysRevX.9.021003} {\bibfield
  {journal} {\bibinfo  {journal} {Phys. Rev. X}\ }\textbf {\bibinfo {volume}
  {9}},\ \bibinfo {pages} {021003} (\bibinfo {year} {2019})}\BibitemShut
  {NoStop}%
\bibitem [{\citenamefont {Krajnik}\ and\ \citenamefont
  {Prosen}(2020)}]{Krajnik2020}%
  \BibitemOpen
  \bibfield  {author} {\bibinfo {author} {\bibfnamefont {{\v{Z}}.}~\bibnamefont
  {Krajnik}}\ and\ \bibinfo {author} {\bibfnamefont {T.}~\bibnamefont
  {Prosen}},\ }\bibfield  {title} {\bibinfo {title} {Kardar--parisi--zhang
  physics in integrable rotationally symmetric dynamics on discrete space--time
  lattice},\ }\href {https://doi.org/10.1007/s10955-020-02523-1} {\bibfield
  {journal} {\bibinfo  {journal} {J. Stat. Phys.}\ }\textbf {\bibinfo {volume}
  {179}},\ \bibinfo {pages} {110} (\bibinfo {year} {2020})}\BibitemShut
  {NoStop}%
\bibitem [{\citenamefont {Klobas}\ and\ \citenamefont
  {Prosen}(2020)}]{Klobas2020}%
  \BibitemOpen
  \bibfield  {author} {\bibinfo {author} {\bibfnamefont {K.}~\bibnamefont
  {Klobas}}\ and\ \bibinfo {author} {\bibfnamefont {T.}~\bibnamefont
  {Prosen}},\ }\href {https://arxiv.org/abs/2004.01671} {\bibinfo {title}
  {Space-like dynamics in a reversible cellular automaton}} (\bibinfo {year}
  {2020}),\ \Eprint {https://arxiv.org/abs/2004.01671} {arXiv:2004.01671}
  \BibitemShut {NoStop}%
\bibitem [{\citenamefont {Wolfram}(1983)}]{Wolfram1983}%
  \BibitemOpen
  \bibfield  {author} {\bibinfo {author} {\bibfnamefont {S.}~\bibnamefont
  {Wolfram}},\ }\bibfield  {title} {\bibinfo {title} {Statistical mechanics of
  cellular automata},\ }\href {https://doi.org/10.1103/RevModPhys.55.601}
  {\bibfield  {journal} {\bibinfo  {journal} {Rev. Mod. Phys.}\ }\textbf
  {\bibinfo {volume} {55}},\ \bibinfo {pages} {601} (\bibinfo {year}
  {1983})}\BibitemShut {NoStop}%
\bibitem [{\citenamefont {Ilachinski}(2001)}]{Ilachinski2001}%
  \BibitemOpen
  \bibfield  {author} {\bibinfo {author} {\bibfnamefont {A.}~\bibnamefont
  {Ilachinski}},\ }\href@noop {} {\emph {\bibinfo {title} {Cellular automata: a
  discrete universe}}}\ (\bibinfo  {publisher} {World Scientific Publishing
  Company},\ \bibinfo {year} {2001})\BibitemShut {NoStop}%
\bibitem [{\citenamefont {Takesue}(1987)}]{Takesue1987}%
  \BibitemOpen
  \bibfield  {author} {\bibinfo {author} {\bibfnamefont {S.}~\bibnamefont
  {Takesue}},\ }\bibfield  {title} {\bibinfo {title} {Reversible cellular
  automata and statistical mechanics},\ }\href
  {https://doi.org/10.1103/PhysRevLett.59.2499} {\bibfield  {journal} {\bibinfo
   {journal} {Phys. Rev. Lett.}\ }\textbf {\bibinfo {volume} {59}},\ \bibinfo
  {pages} {2499} (\bibinfo {year} {1987})}\BibitemShut {NoStop}%
\bibitem [{\citenamefont {Prosen}\ and\ \citenamefont
  {Mej{\'\i}a-Monasterio}(2016)}]{Prosen2016}%
  \BibitemOpen
  \bibfield  {author} {\bibinfo {author} {\bibfnamefont {T.}~\bibnamefont
  {Prosen}}\ and\ \bibinfo {author} {\bibfnamefont {C.}~\bibnamefont
  {Mej{\'\i}a-Monasterio}},\ }\bibfield  {title} {\bibinfo {title}
  {{Integrability of a deterministic cellular automaton driven by stochastic
  boundaries}},\ }\href {http://stacks.iop.org/1751-8121/49/i=18/a=185003}
  {\bibfield  {journal} {\bibinfo  {journal} {J. Phys. A: Math. Theor.}\
  }\textbf {\bibinfo {volume} {49}},\ \bibinfo {pages} {185003} (\bibinfo
  {year} {2016})}\BibitemShut {NoStop}%
\bibitem [{\citenamefont {Inoue}\ and\ \citenamefont
  {Takesue}(2018)}]{Inoue2018}%
  \BibitemOpen
  \bibfield  {author} {\bibinfo {author} {\bibfnamefont {A.}~\bibnamefont
  {Inoue}}\ and\ \bibinfo {author} {\bibfnamefont {S.}~\bibnamefont
  {Takesue}},\ }\bibfield  {title} {\bibinfo {title} {Two extensions of exact
  nonequilibrium steady states of a boundary-driven cellular automaton},\
  }\href {https://doi.org/10.1088/1751-8121/aadc29} {\bibfield  {journal}
  {\bibinfo  {journal} {J. Phys. A: Math. Theor.}\ }\textbf {\bibinfo {volume}
  {51}},\ \bibinfo {pages} {425001} (\bibinfo {year} {2018})}\BibitemShut
  {NoStop}%
\bibitem [{\citenamefont {Prosen}\ and\ \citenamefont
  {Bu\v{c}a}(2017)}]{Prosen2017}%
  \BibitemOpen
  \bibfield  {author} {\bibinfo {author} {\bibfnamefont {T.}~\bibnamefont
  {Prosen}}\ and\ \bibinfo {author} {\bibfnamefont {B.}~\bibnamefont
  {Bu\v{c}a}},\ }\bibfield  {title} {\bibinfo {title} {{Exact matrix product
  decay modes of a boundary driven cellular automaton}},\ }\href
  {http://stacks.iop.org/1751-8121/50/i=39/a=395002} {\bibfield  {journal}
  {\bibinfo  {journal} {J. Phys. A: Math. Theor.}\ }\textbf {\bibinfo {volume}
  {50}},\ \bibinfo {pages} {395002} (\bibinfo {year} {2017})}\BibitemShut
  {NoStop}%
\bibitem [{\citenamefont {{Bu{\v{c}}a}}\ \emph {et~al.}(2019)\citenamefont
  {{Bu{\v{c}}a}}, \citenamefont {{Garrahan}}, \citenamefont {{Prosen}},\ and\
  \citenamefont {{Vanicat}}}]{Buca2019}%
  \BibitemOpen
  \bibfield  {author} {\bibinfo {author} {\bibfnamefont {B.}~\bibnamefont
  {{Bu{\v{c}}a}}}, \bibinfo {author} {\bibfnamefont {J.~P.}\ \bibnamefont
  {{Garrahan}}}, \bibinfo {author} {\bibfnamefont {T.}~\bibnamefont
  {{Prosen}}},\ and\ \bibinfo {author} {\bibfnamefont {M.}~\bibnamefont
  {{Vanicat}}},\ }\bibfield  {title} {\bibinfo {title} {Exact large deviation
  statistics and trajectory phase transition of a deterministic boundary driven
  cellular automaton},\ }\href {https://doi.org/10.1103/PhysRevE.100.020103}
  {\bibfield  {journal} {\bibinfo  {journal} {Phys. Rev. E}\ }\textbf {\bibinfo
  {volume} {100}},\ \bibinfo {pages} {020103} (\bibinfo {year}
  {2019})}\BibitemShut {NoStop}%
\bibitem [{\citenamefont {Friedman}\ \emph {et~al.}(2019)\citenamefont
  {Friedman}, \citenamefont {Gopalakrishnan},\ and\ \citenamefont
  {Vasseur}}]{Friedman2019}%
  \BibitemOpen
  \bibfield  {author} {\bibinfo {author} {\bibfnamefont {A.~J.}\ \bibnamefont
  {Friedman}}, \bibinfo {author} {\bibfnamefont {S.}~\bibnamefont
  {Gopalakrishnan}},\ and\ \bibinfo {author} {\bibfnamefont {R.}~\bibnamefont
  {Vasseur}},\ }\bibfield  {title} {\bibinfo {title} {Integrable many-body
  quantum floquet-thouless pumps},\ }\href
  {https://doi.org/10.1103/PhysRevLett.123.170603} {\bibfield  {journal}
  {\bibinfo  {journal} {Phys. Rev. Lett.}\ }\textbf {\bibinfo {volume} {123}},\
  \bibinfo {pages} {170603} (\bibinfo {year} {2019})}\BibitemShut {NoStop}%
\bibitem [{\citenamefont {Gopalakrishnan}(2018)}]{Gopalakrishnan2018}%
  \BibitemOpen
  \bibfield  {author} {\bibinfo {author} {\bibfnamefont {S.}~\bibnamefont
  {Gopalakrishnan}},\ }\bibfield  {title} {\bibinfo {title} {Operator growth
  and eigenstate entanglement in an interacting integrable floquet system},\
  }\href {https://doi.org/10.1103/PhysRevB.98.060302} {\bibfield  {journal}
  {\bibinfo  {journal} {Phys. Rev. B}\ }\textbf {\bibinfo {volume} {98}},\
  \bibinfo {pages} {060302} (\bibinfo {year} {2018})}\BibitemShut {NoStop}%
\bibitem [{\citenamefont {Gopalakrishnan}\ \emph {et~al.}(2018)\citenamefont
  {Gopalakrishnan}, \citenamefont {Huse}, \citenamefont {Khemani},\ and\
  \citenamefont {Vasseur}}]{Gopalakrishnan2018b}%
  \BibitemOpen
  \bibfield  {author} {\bibinfo {author} {\bibfnamefont {S.}~\bibnamefont
  {Gopalakrishnan}}, \bibinfo {author} {\bibfnamefont {D.~A.}\ \bibnamefont
  {Huse}}, \bibinfo {author} {\bibfnamefont {V.}~\bibnamefont {Khemani}},\ and\
  \bibinfo {author} {\bibfnamefont {R.}~\bibnamefont {Vasseur}},\ }\bibfield
  {title} {\bibinfo {title} {Hydrodynamics of operator spreading and
  quasiparticle diffusion in interacting integrable systems},\ }\href
  {https://doi.org/10.1103/PhysRevB.98.220303} {\bibfield  {journal} {\bibinfo
  {journal} {Phys. Rev. B}\ }\textbf {\bibinfo {volume} {98}},\ \bibinfo
  {pages} {220303} (\bibinfo {year} {2018})}\BibitemShut {NoStop}%
\bibitem [{\citenamefont {Klobas}\ \emph
  {et~al.}(2019{\natexlab{a}})\citenamefont {Klobas}, \citenamefont {Medenjak},
  \citenamefont {Prosen},\ and\ \citenamefont {Vanicat}}]{Klobas2019}%
  \BibitemOpen
  \bibfield  {author} {\bibinfo {author} {\bibfnamefont {K.}~\bibnamefont
  {Klobas}}, \bibinfo {author} {\bibfnamefont {M.}~\bibnamefont {Medenjak}},
  \bibinfo {author} {\bibfnamefont {T.}~\bibnamefont {Prosen}},\ and\ \bibinfo
  {author} {\bibfnamefont {M.}~\bibnamefont {Vanicat}},\ }\bibfield  {title}
  {\bibinfo {title} {Time-dependent matrix product ansatz for interacting
  reversible dynamics},\ }\href {https://doi.org/10.1007/s00220-019-03494-5}
  {\bibfield  {journal} {\bibinfo  {journal} {Commun. Math. Phys.}\ }\textbf
  {\bibinfo {volume} {371}},\ \bibinfo {pages} {651} (\bibinfo {year}
  {2019}{\natexlab{a}})}\BibitemShut {NoStop}%
\bibitem [{\citenamefont {Klobas}\ \emph
  {et~al.}(2019{\natexlab{b}})\citenamefont {Klobas}, \citenamefont {Vanicat},
  \citenamefont {Garrahan},\ and\ \citenamefont {Prosen}}]{Klobas2019b}%
  \BibitemOpen
  \bibfield  {author} {\bibinfo {author} {\bibfnamefont {K.}~\bibnamefont
  {Klobas}}, \bibinfo {author} {\bibfnamefont {M.}~\bibnamefont {Vanicat}},
  \bibinfo {author} {\bibfnamefont {J.~P.}\ \bibnamefont {Garrahan}},\ and\
  \bibinfo {author} {\bibfnamefont {T.}~\bibnamefont {Prosen}},\ }\href
  {https://arxiv.org/abs/1912.09742} {\bibinfo {title} {Matrix product state of
  multi-time correlations}} (\bibinfo {year} {2019}{\natexlab{b}}),\ \Eprint
  {https://arxiv.org/abs/1912.09742} {arXiv:1912.09742} \BibitemShut {NoStop}%
\bibitem [{\citenamefont {Alba}\ \emph {et~al.}(2019)\citenamefont {Alba},
  \citenamefont {Dubail},\ and\ \citenamefont {Medenjak}}]{Alba2019}%
  \BibitemOpen
  \bibfield  {author} {\bibinfo {author} {\bibfnamefont {V.}~\bibnamefont
  {Alba}}, \bibinfo {author} {\bibfnamefont {J.}~\bibnamefont {Dubail}},\ and\
  \bibinfo {author} {\bibfnamefont {M.}~\bibnamefont {Medenjak}},\ }\bibfield
  {title} {\bibinfo {title} {Operator entanglement in interacting integrable
  quantum systems: the case of the rule 54 chain},\ }\href
  {https://doi.org/10.1103/PhysRevLett.122.250603} {\bibfield  {journal}
  {\bibinfo  {journal} {Phys. Rev. Lett.}\ }\textbf {\bibinfo {volume} {122}},\
  \bibinfo {pages} {250603} (\bibinfo {year} {2019})}\BibitemShut {NoStop}%
\bibitem [{\citenamefont {Alba}(2020)}]{Alba2020}%
  \BibitemOpen
  \bibfield  {author} {\bibinfo {author} {\bibfnamefont {V.}~\bibnamefont
  {Alba}},\ }\href {https://arxiv.org/abs/2006.02788} {\bibinfo {title}
  {Diffusion and operator entanglement spreading}} (\bibinfo {year} {2020}),\
  \Eprint {https://arxiv.org/abs/2006.02788} {arXiv:2006.02788} \BibitemShut
  {NoStop}%
\bibitem [{\citenamefont {Fendley}\ \emph {et~al.}(2004)\citenamefont
  {Fendley}, \citenamefont {Sengupta},\ and\ \citenamefont
  {Sachdev}}]{Fendley2004}%
  \BibitemOpen
  \bibfield  {author} {\bibinfo {author} {\bibfnamefont {P.}~\bibnamefont
  {Fendley}}, \bibinfo {author} {\bibfnamefont {K.}~\bibnamefont {Sengupta}},\
  and\ \bibinfo {author} {\bibfnamefont {S.}~\bibnamefont {Sachdev}},\
  }\bibfield  {title} {\bibinfo {title} {Competing density-wave orders in a
  one-dimensional hard-boson model},\ }\href
  {https://doi.org/10.1103/PhysRevB.69.075106} {\bibfield  {journal} {\bibinfo
  {journal} {Phys. Rev. B}\ }\textbf {\bibinfo {volume} {69}},\ \bibinfo
  {pages} {075106} (\bibinfo {year} {2004})}\BibitemShut {NoStop}%
\bibitem [{\citenamefont {Korepin}\ \emph {et~al.}(1997)\citenamefont
  {Korepin}, \citenamefont {Bogoliubov},\ and\ \citenamefont
  {Izergin}}]{korepin1997quantum}%
  \BibitemOpen
  \bibfield  {author} {\bibinfo {author} {\bibfnamefont {V.~E.}\ \bibnamefont
  {Korepin}}, \bibinfo {author} {\bibfnamefont {N.~M.}\ \bibnamefont
  {Bogoliubov}},\ and\ \bibinfo {author} {\bibfnamefont {A.~G.}\ \bibnamefont
  {Izergin}},\ }\href@noop {} {\emph {\bibinfo {title} {Quantum inverse
  scattering method and correlation functions}}},\ Vol.~\bibinfo {volume} {3}\
  (\bibinfo  {publisher} {Cambridge university press},\ \bibinfo {year}
  {1997})\BibitemShut {NoStop}%
\bibitem [{\citenamefont {Sutherland}(2004)}]{sutherland2004beautiful}%
  \BibitemOpen
  \bibfield  {author} {\bibinfo {author} {\bibfnamefont {B.}~\bibnamefont
  {Sutherland}},\ }\href@noop {} {\emph {\bibinfo {title} {Beautiful models: 70
  years of exactly solved quantum many-body problems}}}\ (\bibinfo  {publisher}
  {World Scientific Publishing Company},\ \bibinfo {year} {2004})\BibitemShut
  {NoStop}%
\bibitem [{\citenamefont {Baxter}(2016)}]{baxter2016exactly}%
  \BibitemOpen
  \bibfield  {author} {\bibinfo {author} {\bibfnamefont {R.~J.}\ \bibnamefont
  {Baxter}},\ }\href@noop {} {\emph {\bibinfo {title} {Exactly solved models in
  statistical mechanics}}}\ (\bibinfo  {publisher} {Elsevier},\ \bibinfo {year}
  {2016})\BibitemShut {NoStop}%
\bibitem [{\citenamefont {Iadecola}\ and\ \citenamefont
  {Vijay}(2020)}]{Iadecola2020}%
  \BibitemOpen
  \bibfield  {author} {\bibinfo {author} {\bibfnamefont {T.}~\bibnamefont
  {Iadecola}}\ and\ \bibinfo {author} {\bibfnamefont {S.}~\bibnamefont
  {Vijay}},\ }\href@noop {} {\bibinfo {title} {Nonergodic quantum dynamics from
  deformations of classical cellular automata}} (\bibinfo {year} {2020}),\
  \Eprint {https://arxiv.org/abs/2006.02440} {arXiv:2006.02440} \BibitemShut
  {NoStop}%
\end{thebibliography}%

\appendix

\section{\label{sec:MEstate} MPS for maximum entropy state}
When $\xi=\omega=1$ the MPS representation simplifies. In particular, it can be equivalently expressed as
\begin{equation}\label{eq:equivalentReducedMPS}
    \left.\tr(\vW_1\vV_2\cdots\vV_N)\right|_{\xi,\omega\to1}
    =\tr(\bm{W}_1\bm{W}_2\cdots\bm{W}_N),
\end{equation}
where $W_0$ and $W_1$ are the following $2\times 2$ matrices
\begin{equation}
 W_0=\begin{bmatrix}
        1&1\\
        0&0
    \end{bmatrix},\qquad
    W_1=\begin{bmatrix}
        0&0\\
        1&0
    \end{bmatrix}.
\end{equation}
To see that the two representations are equivalent, we first introduce $4\times 2$ and $2\times 4$ matrices $Q$ and $R$
\begin{equation}
    Q=\begin{bmatrix}
    1&0&1&1\\
    0&1&0&0
    \end{bmatrix},\qquad
    R=\begin{bmatrix}
    1&0\\
    0&1\\
    1&0\\
    -1&0
    \end{bmatrix},
\end{equation}
that map $\WV_n$ into a set of $2\times 2$ matrices $\{W_{n}\}_{n=0,1}$,
\begin{equation}
    W_n=\left.Q \W_n R\right|_{\xi,\omega\to 1}=\left.Q \V_n R\right|_{\xi,\omega\to 1}.
\end{equation}
Therefore, to prove the equivalence, we have to show that the matrix product $RQ$ can be inserted between every pair of matrices on the left-hand side of~\eqref{eq:equivalentReducedMPS}. This follows from the following two relations that hold for any three-site configuration~$(n_1,n_2,n_3)$,
\begin{equation}
\begin{aligned}
\left.\W_{n_1}\V_{n_2}RQ\W_{n_3}\right|_{\xi,\omega\to 1}
&=\left.\W_{n_1}\V_{n_2}\W_{n_3}\right|_{\xi,\omega\to 1},\\
\left.\W_{n_1}R Q\V_{n_2}RQ\W_{n_3}\right|_{\xi,\omega\to 1}
&=\left.\W_{n_1}RQ\V_{n_2}\W_{n_3}\right|_{\xi,\omega\to 1},\\
\end{aligned}
\end{equation}
and the cyclic property of trace.

The stationarity of the right-hand side of Eq.~\eqref{eq:equivalentReducedMPS} can be directly demonstrated by an analogue of the three-site algebraic relation~\eqref{eq:3siteAlgebra}, which in this case trivializes,
\begin{equation}\label{eq:trivial3siteAlgebra}
    \U\left(\bm{W}_1\bm{W}_2\bm{W}_3\right)=\bm{W}_1\bm{W}_2\bm{W}_3.
\end{equation}
The reduced MPS can be understood as the \emph{maximum entropy state} in the restricted sector: every configuration is equally likely, as long as there are no pairs of consecutive
$1$.

\section{\label{sec:partition-function-equivalence} Equivalence of the two forms of the partition sum}

To prove the equivalence of the partition functions in Eqs.~\eqref{eq:partition-function2} and \eqref{eq:partition-function}, we first express the product of transfer matrices as a recursion relation of the form,
\begin{equation}
    T^{K} = T T^{K - 1},
\end{equation}
with matrix elements, denoted by $T^{K}_{jk}$, given by
\begin{equation}\label{eq:transfer-matrix-elements}
    T^{K}_{jk} = \sum_{i = 1}^{4} T^{}_{ji} T_{ik}^{K - 1}.
\end{equation}
where we have introduced the parameter $K$, defined as $2K = N$, to ease the notation. Substituting this parametrization into Eq.~\eqref{eq:partition-function2} admits the following expression for the partition function,
\begin{equation}
    Z = \sum_{i = 1}^{4} T^{K}_{ii}.
\end{equation}

Before searching for a solution to the system of equations in~\eqref{eq:transfer-matrix-elements}, we note that there is significant redundancy in the components of the transfer matrix which we wish to eliminate. Indeed, one can show that the elements of $T^{K}$ can be succinctly written in terms of just four free recursive parameters,
\begin{equation}\label{eq:transfer-matrix-parametrization}
\begin{aligned}
    T^{K}_{11} & = T^{K}_{22}, \\
    T^{K}_{12} & = T^{K}_{12}, \\
    T^{K}_{13} & = T^{K}_{32} + \xi T^{K}_{42}, \\
    T^{K}_{14} & = \xi T^{K}_{32} + \omega T^{K}_{42}, \\
    T^{K}_{21} & = T^{K}_{32} + \xi T^{K}_{42} + \omega T^{K}_{12}, \\
    T^{K}_{22} & = T^{K}_{22}, \\
    T^{K}_{23} & = T^{K}_{12} + \xi T^{K}_{32} + \omega T^{K}_{42}, \\
    T^{K}_{24} & = \xi T^{K}_{12} + \omega T^{K}_{32} + \xi \omega T^{K}_{42}, \\
\end{aligned} \qquad
\begin{aligned}
    T^{K}_{31} & = T^{K}_{12} + \omega T^{K}_{42}, \\
    T^{K}_{32} & = T^{K}_{32}, \\
    T^{K}_{33} & = T^{K}_{22}, \\
    T^{K}_{34} & = \xi T^{K}_{42} + \omega T^{K}_{12}, \\
    T^{K}_{41} & = T^{K}_{32}, \\
    T^{K}_{42} & = T^{K}_{42}, \\
    T^{K}_{43} & = T^{K}_{12}, \\
    T^{K}_{44} & = T^{K}_{22} - T^{K}_{42}.
\end{aligned}
\end{equation}
This parametrization reduces Eq.~\eqref{eq:transfer-matrix-elements} into the remaining four relations,
\begin{equation}\label{eq:transfer-matrix-solutions}
\begin{aligned}
    T^{K}_{12} & = T^{K - 1}_{12} + \xi T^{K - 1}_{32} + \omega T^{K - 1}_{42}, \\
    T^{K}_{22} & = \xi T^{K - 1}_{12} + T^{K - 1}_{22} + \omega T^{K - 1}_{32} + \xi \omega T^{K - 1}_{42}, \\
    T^{K}_{32} & = \omega T^{K - 1}_{12} + T^{K - 1}_{32} + \xi T^{K - 1}_{42}, \\
    T^{K}_{42} & = T^{K - 1}_{22}.
\end{aligned}
\end{equation}
Combining~\eqref{eq:transfer-matrix-parametrization} and~\eqref{eq:transfer-matrix-solutions}
provides an expression for the partition function in terms of one recursive parameter,
\begin{equation}
    Z = 4 T^{K}_{22} - T^{K - 1}_{22},
\end{equation}
for which, Eq.~\eqref{eq:transfer-matrix-solutions} can be rewritten as a higher order recurrence
relation,
\begin{equation}
\begin{aligned}
    T^{K}_{22} & = 3 T^{K - 1}_{22} + (2 \xi \omega - 3) T^{K - 2}_{22} + (1 - \xi \omega) T^{K - 3}_{22} \\ & \qquad + (\xi^{3} + \omega^{3} - \xi^{2} \omega^{2} - \xi \omega) T^{K - 4}_{22}.
\end{aligned}
\end{equation}

To relate this expression for the partition function to Eq.~\eqref{eq:partition-function} it suffices to find a combinatoric form for $T^{K}_{22}$,
\begin{equation}
    T^{K}_{22} = \sum_{\{Q\}} C^{K}_{Q} \xi^{Q^{+}} \omega^{Q^{-}},
\end{equation}
where $C^{K}_{Q} = C(K, Q^{+}, Q^{-})$ is some combinatoric factor to be determined and the set $\{Q\}$ the set of tuples of positive and negative quasiparticle numbers satisfying the constraints in Eqs.~\eqref{eq:quasiparticle-constraint} and \eqref{eq:quasiparticle-size-constraint}. With a little work, one can show that the combinatoric term is given by
\begin{equation}
    C^{K}_{Q} = \binom{K - \frac{1}{3} Q^{+} - \frac{2}{3} Q^{-}}{Q^{+}} \binom{K - \frac{1}{3} Q^{-} - \frac{2}{3} Q^{+}}{Q^{-}}.
\end{equation}
The partition function can then be rewritten as
\begin{equation}\label{eq:transfer-matrix-combinatorial}
    Z = \sum_{\{Q\}} \big(4 C^{K}_{Q} - C^{K - 1}_{Q}\big) \xi^{Q^{+}} \omega^{Q^{-}},
\end{equation}
where to combine summations we have used the property that the binomial coefficients vanish when Eq.~\eqref{eq:quasiparticle-size-constraint} is not satisfied. Utilising the binomial identity $\binom{n - 1}{k} = \frac{n - k}{n} \binom{n}{k}$, we can express $C^{K - 1}_{Q}$ in terms of $C^{K}_{Q}$, specifically,
\begin{equation}
    C^{K - 1}_{Q} = \frac{\big(K - \frac{2}{3} Q^{+} - \frac{4}{3} Q^{-}\big)\big(K - \frac{2}{3} Q^{-} - \frac{4}{3} Q^{+}\big)}{\big(K - \frac{1}{3} Q^{+} - \frac{2}{3} Q^{-}\big)\big(K - \frac{1}{3} Q^{-} - \frac{2}{3} Q^{+}\big)} C^{K}_{Q}.
\end{equation}
From here, with a simple substitution, we immediately see that this expression for the partition function is exactly equivalent to that in Eq.~\eqref{eq:quasiparticle-entropy}, where the combinatorial coefficients follow directly as
\begin{equation}
    4 C^{K}_{Q} - C^{K - 1}_{Q} = \frac{1}{m_{Q}} C^{K}_{Q} = \Omega_{Q}.
\end{equation}

\end{document}